\documentclass[article,nojss]{jss}

\ifdefined\Plainauthor
\else
  \usepackage[authoryear,round]{natbib}
  \usepackage{graphicx,fancyvrb,hyperref}
  \newcommand{\Plainauthor}[1]{}
  \newcommand{\Plaintitle}[1]{}
  \newcommand{\Shorttitle}[1]{}
  \newcommand{\Plainkeywords}[1]{}
  \newcommand{\And}{\and}
  \newcommand{\pkg}[1]{\textbf{#1}}
  \newcommand{\proglang}[1]{\textsf{#1}}
  \newcommand{\code}[1]{\texttt{#1}}
  \newcommand{\email}[1]{\href{mailto:#1}{\texttt{#1}}}
  \newcommand{\doi}[1]{\href{https://doi.org/#1}{\texttt{doi:#1}}}
  \providecommand{\E}{\mathsf{E}}
  \newcommand{\Abstract}[1]{\AtBeginDocument{\begin{abstract}#1\end{abstract}}}
  \newcommand{\Keywords}[1]{\AtBeginDocument{\noindent\textbf{Keywords:} #1\par}}
  \newcommand{\Address}[1]{\AtEndDocument{\section*{Affiliations}#1}}
  \newenvironment{CodeChunk}{}{}
  \DefineVerbatimEnvironment{Code}{Verbatim}{}
  \DefineVerbatimEnvironment{CodeInput}{Verbatim}{}
  \DefineVerbatimEnvironment{CodeOutput}{Verbatim}{}
  \date{}
  \AtBeginDocument{\maketitle}
\fi

\usepackage{amsmath,amssymb,booktabs,longtable,array,needspace}
\renewcommand{\E}{\mathbb{E}}
\newcommand{\Prb}{\mathbb{P}}
\author{Sunday A. Adetunji\\Oregon State University
 \And Oluwatimileyin A. Adetunji\\Bowen University}
\Plainauthor{Sunday A. Adetunji, Oluwatimileyin A. Adetunji}
\title{\pkg{coherentSurv}: Coherent Survival Modelling and Dynamic Forecasting for Recurrent and Terminal Events in \proglang{R} and \proglang{Python}}
\Plaintitle{coherentSurv: Coherent Survival Modelling and Dynamic Forecasting for Recurrent and Terminal Events in R and Python}
\Shorttitle{Coherent Survival Forecasting in \proglang{R} and \proglang{Python}}
\Abstract{Researchers studying recurrent events and terminal outcomes need software that connects incomplete event histories to interpretable forecasts and uncertainty estimates. We present \pkg{coherentSurv}, an \proglang{R} package for fitting history-dependent recurrent and terminal event models, predicting future event burden and terminal-event risk, and assessing sensitivity to incomplete observation. Users supply event histories, covariates and observation information; the package returns fitted models, dynamic forecasts, uncertainty summaries and diagnostic measures under explicitly stated assumptions.

The package integrates over event histories compatible with tied dates or observation windows and supports several frailty distributions for between-subject heterogeneity. A shared probability model generates recurrent-event counts, terminal-event probabilities and restricted mean time alive. Cross-fitted targeting provides influence-function diagnostics and asymptotic inference in supported missing-at-random and right-censoring settings. Separate sensitivity analyses provide bounds under specified recording restrictions. A baseline binary-treatment interface estimates standardized treatment contrasts under consistency, conditional exchangeability, positivity and independent censoring.

A compiled \proglang{C++}17 core supports an \proglang{R} interface and \proglang{Python} access to the same implementation through \pkg{rpy2} and \code{Rscript}. Independent numerical checks, computational benchmarks, simulation experiments and reproducible examples document accuracy, performance and inferential limitations. Outputs retain observation assumptions, convergence information and numerical diagnostics. The package, tutorials and replication materials are publicly available, supporting transparent analysis across clinical and other event-history applications.}
\Plainkeywords{recurrent events, terminal event, frailty, dynamic prediction, coarsened data, partial identification, R, Python, C++}
\Address{Sunday A. Adetunji\\Department of Biostatistics and Epidemiology\\College of Health, Oregon State University\\Corvallis, OR, USA\\Corresponding author: \email{sundayadetunjisa@gmail.com}
\medskip\\Oluwatimileyin A. Adetunji\\Department of Economics\\College of Management and Social Sciences, Bowen University\\Iwo, Osun State, Nigeria\\E-mail: \email{eniola16.to@gmail.com}}
\begin{document}
% Numerical tables and figures are generated by replication_script.R.

\section{Introduction}
\label{sec:intro}
A recurrent-event forecast can mean an expected event count, a probability of an event, or a probability of remaining alive. These quantities are related, but their units and eligibility rules differ. A patient may be readmitted repeatedly before death; an equipment unit may undergo several repairs before irrecoverable failure. A useful implementation must carry these distinctions from data preparation through estimation, prediction and validation. It must also distinguish events that did not occur from intervals in which events were not recorded.

The methodological framework of \citet{Adetunji2026} constructs prognostic targets from a common complete-history law and separates the assumptions needed for likelihood inference, missing-history sensitivity analysis and prediction-set calibration. The present paper describes its software implementation. The primary contribution is an auditable computational path: declared event eligibility and observation restrictions enter a shared likelihood; one retained law generates multiple forecast targets; independent numerical routines report their own assumptions and diagnostics. The software paper complements the methodological manuscript rather than claiming that all of its theoretical extensions are already available as general-purpose routines.

The package addresses three practical sources of implementation error. First, an observed terminal event changes every subsequent risk set. Second, a coarsened history is a set of compatible paths; choosing a convenient representative path can change the likelihood and forecast. Third, finite-sample prediction coverage, parameter confidence coverage and numerical precision are different properties. A common function name or a successful optimizer does not make their guarantees interchangeable. The interfaces expose these distinctions so that analysts can decide which calculation answers their question.

The core architecture introduced in version 0.10.0 integrates five extensions: automatic observation fibres, sparse propagation and large-support bounds, targeting through the primary modeling API, evolving forecast covariates, and alternative frailty distributions with diagnostics. All data-processing and high-level methods are available in \proglang{R}; the likelihood, coarsened mixture, propagation and core finite-support operations are implemented through a separate \proglang{C++17} library. We report deterministic numerical agreement, measured computational performance and five case studies. Version 0.10.1 adds explicit data-regime tags, survival and competing-risk wrappers, and audit export. Its causal revision adds a separate treatment-weighted targeting engine implementing Supplement S12 of the methodological manuscript; existing prognostic numerical kernels are retained.

In an acute-care cohort of women with uterine, tubo-ovarian or cervical malignancies, Cox regression was complemented by 365-day restricted mean survival time and a hierarchical Bayesian piecewise-exponential model to describe one-year all-cause prognosis after the index hospitalization \citep{AdetunjiOyewusi2026}.

\section{Related software}\label{sec:related}
Several established packages provide essential recurrent-event and multistate tools. \pkg{frailtypack} fits shared, joint and nested frailty models and includes dynamic prediction for longitudinal, recurrent and terminal outcomes \citep{Rondeau2012,Krol2017}. \pkg{mstate} organizes competing-risk and multistate data and supports transition-specific estimation and prediction \citep{deWreede2011}. \pkg{reReg} implements a general scale-change regression framework for recurrent events, including informative terminal events and frailty treatment without a required parametric frailty distribution \citep{Chiou2023}. \pkg{JMbayes2} fits Bayesian joint models linking several longitudinal outcomes and event processes, with prediction and validation utilities \citep{JMbayes2}. These packages have different scientific scopes; a single ranking would conceal the differences.

Table~\ref{tab:related} records the relevant documented capabilities and the corresponding role of \pkg{coherentSurv}. The comparison does not claim that the other packages necessarily mishandle missing data or produce incoherent forecasts. In particular, a correctly specified joint model already supplies a common probability law. The additional interfaces here make legal-path integration, finite-path targeting, recording-sensitivity optimization and explicit numerical checks available in one package, within their stated observation and state models. Version 0.10.0 additionally provides continuous-time targeting for a declared finite-state right-censoring experiment and automatic construction of observed-event timing fibres.

\begin{table}[tbp]\centering\small
\begin{tabular}{>{\raggedright\arraybackslash}p{.15\textwidth}>{\raggedright\arraybackslash}p{.40\textwidth}>{\raggedright\arraybackslash}p{.36\textwidth}}\toprule
Software & Documented emphasis & Relationship to this implementation \\ \midrule
\pkg{frailtypack} & Shared, joint and nested frailty estimation; dynamic prediction \citep{Rondeau2012,Krol2017}. & Established frailty workflow; the present interfaces additionally expose legal-path integration and finite-support sensitivity constraints. \\ 
\pkg{mstate} & Competing-risk and multistate data preparation, estimation and prediction \citep{deWreede2011}. & Transition-oriented workflow; retained count states here encode selected history dependence conditional on frailty. \\ 
\pkg{reReg} & Recurrent-event scale-change regression with possibly informative terminal events \citep{Chiou2023}. & Different regression and frailty assumptions; no universal accuracy or runtime ranking is implied. \\ 
\pkg{JMbayes2} & Bayesian longitudinal and event-time joint models, prediction and validation \citep{JMbayes2}. & Richer longitudinal submodels; the present finite-path and sensitivity calculations have distinct input contracts. \\ 
\pkg{coherentSurv} & Automatic fibres, alternative frailty, finite-path and continuous targeting, evolving covariates and sparse forecasts/bounds. & Explicitly separates working-model forecasts, confidence regions and prediction sets. \\ 
\bottomrule\end{tabular}
\caption{Complementary software scopes. Statements describe documented capabilities; only the companion examples identified in the text were executed here. Absence from this table is not an assertion that another package cannot implement an extension.}\label{tab:related}
\end{table}

For a basic execution comparison, the replication script also fits a joint Weibull gamma model to the public readmission data using \pkg{frailtypack}, and a terminal-event Cox model to the colon trial using \pkg{survival} \citep{survival}. The commands and installed versions are recorded with the outputs. Their likelihoods, baseline specifications and targets differ from the \pkg{coherentSurv} examples; we therefore do not compare their likelihood values as model scores or report an unmatched speed ratio.

\section{Model and target quantities}\label{sec:model}
Let $i=1,\ldots,n$ index independent subjects and $k=1,\ldots,K$ recurrent event types. Write $N_{ik}(t)$ for a counting process, $T_i^D$ for terminal time, $X_i(t)$ for predictable measured covariates and $h_i(t-)$ for a retained history summary. The working intensity, conditional on a mean-one frailty, is
\begin{equation}\label{eq:intensity}
\begin{aligned}
\lambda_{ik}(t\mid Z_i,\mathcal F_{i,t-})&=
Y_{ik}(t)Z_i\lambda_{0k}(t)
\exp\{\beta_k^\top X_i(t)+\gamma_k^\top h_i(t-)\},\\
Z_i&\sim F_\eta,\qquad \E_{F_\eta}(Z_i)=1.
\end{aligned}
\end{equation}
Here $F_\eta$ is a frailty distribution with parameter $\eta$, $Y_{ik}(t)$ is eligibility immediately before $t$, $Z_i$ is mean-one frailty with variance $\phi$, $\lambda_{0k}$ is a piecewise-constant baseline, and $\mathcal F_{i,t-}$ is the complete past. A terminal intensity has its own coefficients. Eligibility is zero after the terminal event; a non-repeatable mark also becomes ineligible after its first occurrence. The package's standard history state consists of capped prior counts. The cap changes the intensity feature, not the number of future events included in a burden reward.

At a surviving landmark $s$ and horizon $\Delta>0$, the target classes are
\begin{equation}\label{eq:targets}
\begin{aligned}
\pi_D(s,\Delta)&=\Prb(T^D\leq s+\Delta\mid T^D>s,\mathcal H_s),\\
\mu_k(s,\Delta)&=\E\{N_k(s+\Delta)-N_k(s)\mid T^D>s,\mathcal H_s\},\\
r(s,\Delta)&=\E\{\min(T^D-s,\Delta)\mid T^D>s,\mathcal H_s\}.
\end{aligned}
\end{equation}
Here $\mathcal H_s$ denotes the available record, $\pi_D$ is terminal risk, $\mu_k$ is expected additional type-$k$ burden, and $r$ is restricted mean time alive. Counts stop at the terminal event. These are prognostic expectations. Interpreting a coefficient as the effect of preventing a recurrence would require additional causal assumptions and a different estimand.

For an exact path, let $d$ count all eligible jumps, $B$ be the sum of log intensities excluding frailty at those jumps, and $H$ be integrated total intensity excluding frailty. For $F_\eta=\mathrm{Gamma}(a,a)$ with $a=1/\phi$, analytic integration yields
\begin{equation}\label{eq:likelihood}
\ell=B+\log\Gamma(a+d)-\log\Gamma(a)+a\log a-(a+d)\log(a+H).
\end{equation}
Here $\ell$ is the path's marginal log likelihood, $\Gamma$ is the Euler gamma function and $a=1/\phi$. The conditional frailty distribution is $\mathrm{Gamma}(a+d,a+H)$. The kernel uses integer-count recurrences and stable logarithmic evaluations to avoid cancellation near $\phi=0$; the zero-variance model is available explicitly. The complete derivations and identification conditions are in the methodological manuscript. They include independent subjects, predictable covariates, non-explosion, a valid observation model and sufficient retained state.

\section{Software architecture and numerical algorithms}\label{sec:architecture}
\begin{figure}[t]
\centering\includegraphics[width=\textwidth]{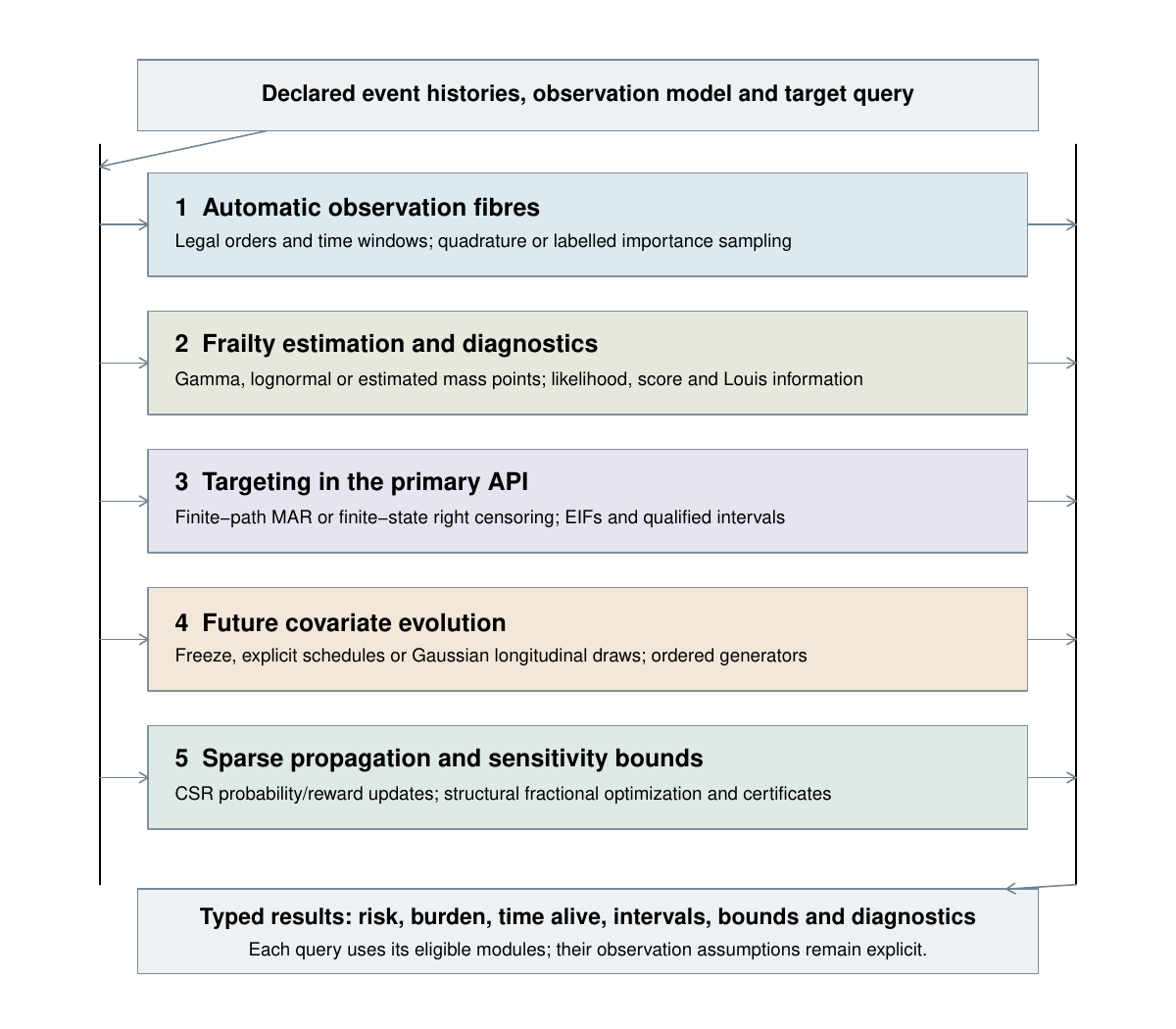}
\caption{Five implemented extensions and their computational dependencies. Arrows transmit data or calculations, not causal effects. Targeting and sensitivity bounds retain distinct observation experiments; every output carries its assumptions and diagnostics.}\label{fig:architecture}
\end{figure}

Figure~\ref{fig:architecture} shows the separation of data contracts, numerical kernels and inferential outputs. The C++ headers do not include R headers. Thin \pkg{Rcpp} bridges translate objects and exceptions \citep{Rcpp}; the R layer implements formula handling, S3 classes, case adapters, diagnostics and reporting. This design permits independent standalone kernel tests and avoids duplicating the numerical method inside plotting or data-import code.

\subsection{Data validation and predictable history}
\code{cs\_as\_trajectory()} requires a subject identifier, interval start and stop, recurrent mark, terminal indicator and declared time unit. Optional monitoring indicators identify unobserved intervals. It checks finite times, chronological grouping, overlap, event coding and absorption. Zero-length recurrence--death ties are not silently jittered. With a declared rounding resolution, the coarsening engine converts dates into event-time windows and integrates over their admissible histories. Exact fitting rejects gaps that have no observation model. Delayed entry is rejected by the current frailty fitting interface because prior selection of frailty would otherwise be omitted.

\code{cs\_build\_history()} evaluates counts and recency at interval start, before the current interval's event. Missing recency means that no earlier event is recorded; it is not replaced by a zero waiting time. The model formula cannot contain the interval endpoint or its event outcome. Subject selection must retain complete histories, and the documented missing-value option removes whole affected subjects rather than isolated rows. These checks address temporal leakage and broken risk sets, but cannot validate the scientific meaning of an input code.

The 0.10.1 interface adds the leading class \code{cs\_trajectory} while retaining \code{surv\_trajectory} and \code{data.frame} inheritance. An optional \code{terminal\_cause} field declares the absorbing cause; it is separate from recurrent marks and censoring. Attached pattern metadata summarize what was recorded. A dataset with no repeated events does not by itself establish a single-event design, and several event labels do not prove a competing-risk structure. Eligibility, absorption and the observation experiment remain analyst declarations.

\begin{samepage}
The executable \code{replication/standards\_mapping.R} examples illustrate a selected CDISC ADaM time-to-event parameter and explicitly defined OMOP event concepts \citep{CDISCTTE,CDISCExample,OMOP54}. They are source-to-schema transformations, not certified adapters. For ADTTE, the analyst selects the analysis parameter, preserves its duration and censoring derivation, and maps \code{CNSR == 0} to an event indicator. Multiple positive censoring codes must not be converted by \code{1 - CNSR}. For OMOP, concept sets, episode rules, person linkage, observation periods and terminal ascertainment must be specified before records become event histories. Database observation is not proof of complete event ascertainment.

\end{samepage}

\subsection{Automatic observation-fibre construction}
With \code{observation = "coarsened"}, the primary fit can call \code{cs\_coarsen\_fibre()} instead of requiring manually expanded paths. The user supplies event marks, windows $(L_j,U_j]$, a follow-up endpoint and the time unit. A declared rounding unit converts a point date into a positive-width window; the program does not infer date precision. Exact point events retain counting mass, whereas interval times carry Lebesgue integration mass. Event counts and marks are assumed recorded completely within this experiment. An unobserved gap with an unknown number of events remains a different observation model.

Let $\mathcal F(o)$ contain chronological marked histories compatible with observed record $o$. The terminal mark must be last, and nonrepeatable marks obey their eligibility rules. The likelihood and its numerical approximations are
\begin{equation}\label{eq:fibre}
 L(\theta;o)=\int_{\mathcal F(o)}L_c(\theta;h)\,d\mu(h),\qquad
 \widehat L(\theta;o)=\sum_{b=1}^{M}w_bL_c(\theta;h_b).
\end{equation}
Here $\theta$ is the parameter vector, $L_c$ the complete-history likelihood, $h$ a legal history, $\mu$ the product integration measure, $h_b$ a generated history, $w_b$ its integration mass and $M$ the number of generated quadrature paths or attempted importance draws. Under sampling, a rejected draw contributes zero and $w_b=1/\{M q(h_b)\}$ for proposal density $q$. Integration weights are not normalized before likelihood calculation. Importance integration is unbiased for the likelihood when its support covers the fibre and the integrand is integrable; the log likelihood is not unbiased.

Same-mark windows require particular care. Labelled permutations can count one physical marked history repeatedly. The engine assigns each successive occurrence to the available same-mark window with earliest upper endpoint. An exchange of two feasible assignments preserves membership in interval windows and produces this canonical order. Thus overlapping same-mark windows are integrated as a union. Complete enumeration is followed by Gaussian quadrature split at window endpoints and baseline discontinuities. Exceeding \code{max\_perms} or \code{max\_paths} triggers seeded importance sampling, with attempted/rejected counts, effective sample size and integration MCSE retained. Enumeration removes permutation error; it does not make a general continuous-time integral analytically exact.

\subsection{Likelihood, scores and legal coarsening}
The low-level \code{cs\_loglik()} takes expanded marked exposures and a design matrix, returning subject log likelihoods, analytic scores and observed information. \code{cs\_fit()} constructs that design, splits intervals at baseline changes and optimizes the marginal likelihood. It reports gradient magnitude, information eigenvalues, optimizer status and parameter-boundary flags. A numerical optimizer's success code is insufficient to declare a regular fit.

For coarsened observations, generated or explicitly supplied legal complete paths and log integration weights enter \code{cs\_coarsen()}. With component log likelihoods $\ell_j$, scores $u_j$ and observed information $I_j$, its normalized path weights $w_j$ give
\begin{equation}\label{eq:louis}
\begin{aligned}
\ell_O&=\log\sum_j \nu_j e^{\ell_j},\qquad
u_O=\sum_jw_ju_j,\qquad
w_j=\frac{\nu_je^{\ell_j}}{\sum_v\nu_ve^{\ell_v}},\\
I_O&=\sum_jw_jI_j-\sum_jw_j(u_j-u_O)(u_j-u_O)^\top.
\end{aligned}
\end{equation}
Here $j$ indexes legal quadrature or importance paths, $\nu_j$ is an integration weight and $O$ denotes the observed record. Log-sum-exp evaluates the mixture stably. Equation~\ref{eq:louis} is the missing-information identity \citep{Louis1982}; averaging component information alone would omit the score-variation subtraction. Integration weights are not normalized before likelihood calculation. Refining observation-fibre quadrature remains necessary when continuous event times are integrated numerically.

\subsection{Ordered propagation and cumulative rewards}
Conditional on one frailty value and a sufficient finite state, the generator $Q$ has nonnegative off-diagonal entries and zero row sums. If $\nu\geq\max_j(-Q_{jj})$, uniformization gives
\begin{equation}\label{eq:uniformization}
e^{Q\Delta}=e^{-\nu\Delta}\sum_{m=0}^{\infty}
\frac{(\nu\Delta)^m}{m!}\left(I+Q/\nu\right)^m.
\end{equation}
Here $\Delta$ is segment duration, $m$ is a Poisson-series index and $I$ is the identity. Uniformization replaces the matrix exponential by Poisson-weighted discrete transitions \citep{Rao2013}. The dense \code{cs\_propagate()} remains an independent small-state reference. The new \code{cs\_generator\_csr()} stores off-diagonal rates as compressed sparse rows. Probability row vectors and integrated rewards are updated by \code{cs\_propagate\_sparse()} without an $S\times S$ intermediate matrix. For $S$ states, $E$ stored transitions and $r$ rewards, each retained Poisson term costs $O(E+Sr)$ work and working storage is linear in edges and states. The total number of terms and duration chunks depends on rates, horizon and tolerance; no unconditional linear-time claim is made.

The implementation propagates real-arithmetic series-truncation bounds through ordered segments. Roundoff is reported separately as a diagnostic, not as interval-arithmetic certification. It uses duration chunking and explicit geometric tail inequalities; it does not claim to implement the distinct Fox--Glynn algorithm \citep{Fox1988}. Term and chunk budgets produce structured errors rather than silently returning an unqualified approximation.

Generators are applied in temporal order across baseline changes. A single frailty component is retained for all segments; averaging transition matrices independently at each segment would change the dependence structure. Frailty integration is performed after each component's ordered propagation. \code{cs\_predict\_paths()} adds another outer mixture over explicitly supplied compatible pre-landmark histories. It rejects future paths at an earlier assessment time. No single frailty component is substituted for a posterior mixture merely because it would be easier to store.

\subsection{Targeting through the primary modeling interface}
The choice \code{targeting = "finite\_path"} in \code{cs\_fit()} exposes the whole-path missing-at-random (MAR) experiment implemented by \code{cs\_target()}. Let $R$ indicate that a finite path $Y$ is recorded, $X$ be an observed stratum, $F(Y)$ a vector of target functions, $m(x)=\E\{F(Y)\mid X=x\}$ and $\pi(x)=\Prb(R=1\mid X=x)$. The influence function is
\begin{equation}\label{eq:eif}
D(O)=m(X)-\psi+\frac{R}{\pi(X)}\{F(Y)-m(X)\},\qquad
\psi=\E\{F(Y)\}.
\end{equation}
Here $O=(X,R,RY)$ is the observed unit and $D$ is a vector. Subjects define folds. Training-fold nuisance estimates feed one common exponential tilt; the resulting mixture law generates every target. The function returns individual EIF values, their means and variances, recording estimates, fold assignments and score convergence. Double-robust consistency and regular efficiency retain their separate nuisance and remainder conditions. Whole-path MAR does not cover arbitrary intermittent hidden events.

The option \code{targeting = "continuous"} fits a finite-state counting-process experiment with explicit right censoring. Let $V_j(t,h)$ be the remaining conditional value of target $j$ at time $t$ and retained state $h$, and define the jump contrast $B_{jk}(t,h)=a_{jk}(t,h)+V_j(t,h+k)-V_j(t,h)$. Its observed-data influence function is
\begin{equation}\label{eq:continuous_eif}
 D_j(O_i)=V_j(0,X_i)-\psi_j+
 \sum_{k=1}^{K+1}\int_0^\tau
 \frac{R_i(t)}{G(t\mid X_i)}B_{jk}(t,H_i(t-))
 \{dN_{ik}(t)-\lambda_{ik}(t)\,dt\}.
\end{equation}
Here $i$ indexes independent subjects, $j$ the finite target vector, $k$ the recurrent marks and terminal mark $K+1$, $X_i$ a baseline stratum, $H_i(t-)$ the sufficient pre-jump state, $R_i(t)$ the observation-at-risk indicator, $G$ conditional censoring survival, $N_{ik}$ the counting process, $\lambda_{ik}$ its full-history intensity, $a_{jk}$ the target's immediate jump reward, $\tau$ the assessment horizon and $\psi_j$ the population target. The jump contrast follows from the backward value equation: a jump earns its immediate reward and changes the remaining expected value. This implementation requires contiguous histories from time zero, conditionally independent right censoring within declared baseline strata, positive $G$, and a sufficient retained state.

Subject-level folds estimate event and censoring nuisances on training subjects. On validation subjects the common fluctuation is
\begin{equation}\label{eq:target_update}
 \lambda_k^{\epsilon}(t,h)=\widehat\lambda_k(t,h)
 \exp\!\left\{\sum_j\epsilon_j\frac{\widehat B_{jk}(t,h)}{\widehat G(t\mid x)}\right\}.
\end{equation}
Here $\widehat\lambda_k$ and $\widehat G$ are fold-trained nuisance estimates, $x$ is the baseline stratum, $\widehat B_{jk}$ is the current value contrast, and $\epsilon_j$ is a targeting coefficient. After each update the algorithm recomputes values, contrasts and individual EIFs. Event nuisances may use pooled or count-loglinear working intensities; censoring hazards are constant within the declared strata in the exported implementation. Time-grid refinement checks numerical interpolation error. Cross-fitting alone does not supply the nuisance product-rate, regular targeting and stochastic equicontinuity conditions required for regular asymptotic inference.

The fitted object retains one fold-stratum mixture law. Dynamic conditional prediction updates its fold weights using only the observed history through the landmark. Population target intervals use centered subject-level EIF multipliers and a maximum-$t$ critical value; bounded targets use a logit transform and nonnegative unbounded burdens use a log transform. Undefined transformed or nonregular results are flagged. These intervals are asymptotic finite-vector approximations. The population EIF is not reused as an individual-history confidence interval, and transformation alone does not establish sparse-cell coverage.

When regular inference is unavailable, observed reward envelopes provide a separate conservative fallback. Suppose $a_j\leq L_{ij}\leq F_{ij}\leq U_{ij}\leq b_j$ for independent identically distributed units $i=1,\ldots,n$ and bounded targets $j=1,\ldots,m$. Then
\begin{equation}\label{eq:outer_envelope}
 \left[\max\{a_j,\overline L_j-r_j\},\ \min\{b_j,\overline U_j+r_j\}\right],\qquad
 r_j=(b_j-a_j)\sqrt{\frac{\log(2m/\alpha)}{2n}}
\end{equation}
jointly cover $\E(F_{ij})$ with probability at least $1-\alpha$. Here $F_{ij}$ is the complete target reward, $L_{ij},U_{ij}$ its observable envelopes, $a_j,b_j$ its known support endpoints, the bars denote sample means and $\alpha$ is the familywise error level. Hoeffding's inequality \citep{Hoeffding1963} bounds each of the $2m$ one-sided deviations by $\alpha/(2m)$; the union bound and $\E L_{ij}\leq\E F_{ij}\leq\E U_{ij}$ establish the claim. Terminal risk uses observed death as its lower envelope and adds censoring before the horizon to its upper envelope. Restricted survival uses observed follow-up as its lower envelope and the horizon as its upper envelope when censored. This is a coarsening outer set, not a sharp sensitivity bound or a substitute EIF confidence interval. It allows informative censoring; uncapped burdens return $[0,\infty)$.

For this unconditional finite-sample statement, request \code{interval = "projection"} throughout the prespecified analysis. Automatic data-dependent switching between approximate Gaussian and projection intervals does not inherit either unconditional finite-sample coverage for the combined procedure or selection-conditional coverage after its branch is chosen. The same qualification applies to the automatic zero-event fallback.

\subsection{Baseline treatment effects}\label{sec:causal}
\code{cs\_causal()} implements the static baseline binary-treatment extension in Supplement S12 of \citet{Adetunji2026}. Let $A\in\{0,1\}$ be treatment (denoted $W$ in that supplement), $X$ baseline confounders, $Y^a$ the potential complete history under assignment $a$, and $F_j$ the recurrence-count, terminal-risk or restricted-survival reward. The interventional mean is
\begin{equation}\label{eq:causal_mean}
 \psi_j(a)=\E\{F_j(Y^a)\}=\E\{V_j^a(0,X)\},\qquad a\in\{0,1\}.
\end{equation}
Here $j$ indexes the outcome coordinate and $V_j^a(0,X)$ is its arm-specific backward value at baseline. Identification requires consistency and no interference, treatment exchangeability conditional on $X$, positive treatment probabilities, and conditionally independent right censoring with positive censor survival. The package checks input support and estimated overlap; it cannot verify exchangeability. Both arms are standardized over the same empirical distribution of $X$.

For independent subject $i$, the estimated arm gradient follows the continuous-time targeting principle of \citet{Rytgaard2022}:
\begin{equation}\label{eq:causal_eif}
 \begin{aligned}
 D_j^a(O_i)={}&V_j^a(0,X_i)-\psi_j(a)\\
 &+\frac{\mathbf{1}(A_i=a)}{e_a(X_i)}
 \sum_{k=1}^{K+1}\int_0^\tau
 \frac{R_i(t)B_{jk}^a(t,H_i(t-))}{G_a(t\mid X_i)}
 \{dN_{ik}(t)-\lambda_k^a(t,H_i(t-))\,dt\}.
 \end{aligned}
\end{equation}
Here $O_i$ is the observed history, $e_a(X_i)=\Prb(A_i=a\mid X_i)$ the arm probability, $R_i(t)$ the observation-at-risk indicator, $G_a$ arm-specific censor survival, $k$ an event mark, $N_{ik}$ its counting process, $H_i(t-)$ the retained state, $\lambda_k^a$ its eligible intensity, $B_{jk}^a$ the backward jump contrast and $\tau$ the fixed horizon. Every subject contributes the baseline-value term in both arms; the event-martingale term uses only the subject's observed arm. Subjects without a record in an arm are therefore not dropped from its standardization population.

The treatment- and censoring-weighted update is
\begin{equation}\label{eq:causal_tilt}
 \lambda_k^{a,\epsilon}(t,h)=\widehat\lambda_k^a(t,h)
 \exp\!\left\{\sum_j\epsilon_j
 \frac{\widehat B_{jk}^a(t,h)}{\widehat e_a(x)\widehat G_a(t\mid x)}\right\}.
\end{equation}
Here $a$ identifies the arm, $h$ the retained history with baseline part $x$, $j$ the target coordinate, $k$ the event mark and $\epsilon_j$ its fluctuation coefficient; hats denote fold-trained nuisances or the current backward contrast. The positive multiplier preserves zero ineligible intensities and terminal absorption. Event, treatment and censoring nuisances are trained outside each subject's fold. The pooled validation score determines the targeting update; the updated law is not itself described as an entirely training-fold fit.

The baseline treatment contrast retains the covariance between arms:
\begin{equation}\label{eq:causal_contrast}
 \widehat\Delta_j=\widehat\psi_j(1)-\widehat\psi_j(0),\qquad
 \widehat D_{\Delta,j,i}=\widehat D_j^1(O_i)-\widehat D_j^0(O_i),\qquad
 \widehat\Sigma_\Delta=\frac{\widehat{\operatorname{Cov}}(\widehat D_{\Delta,i})}{n}.
\end{equation}
Here $n$ is the number of independent subjects, $i$ a subject, $j$ a target, $\Delta_j$ the arm-1 minus arm-0 effect, and $\widehat\Sigma_\Delta$ the covariance matrix of its estimator. \code{cs\_causal\_contrast()} also returns ratios and log ratios, using the delta-method gradient and positive log-scale limits for ratios. A ratio of expected counts is a burden ratio, not a hazard ratio. Gaussian multipliers over the small joint covariance matrix give simultaneous asymptotic intervals for regular coordinates without allocating a draws-by-cohort-size matrix.

The exported implementation uses at most 30 prespecified discrete baseline strata, six recurrent marks and 128 living states. Default event nuisances are pooled or log-linear in retained counts; censor hazards are constant within arm and baseline stratum. Out-of-fold event-rate arrays can vary over the declared time grid. Treatment probabilities may be known by design or supplied as fold-by-stratum estimates. No automatic propensity clipping is used. The retained count cap summarizes state dependence; it does not truncate accumulated recurrence rewards.

Consistency of the augmented estimating function requires correct event intensities or a correct product of treatment probability and censor survival. Correct propensity alone does not repair a misspecified censor model. Efficiency and normal inference additionally require the product-rate and regularity conditions in Supplement S12. Zero-event, nonconverged or zero-variance coordinates retain their estimates and diagnostic flags but receive no regular confidence limits. The prognostic recording-sensitivity and observed-envelope intervals are not substituted for causal treatment bounds. Time-varying treatment, unmeasured treatment-confounding sensitivity and coarsened causal histories require further estimators and are outside this entry point.

\subsection{Large-support sensitivity optimization}
\code{cs\_bounds()} takes a recorded-category distribution, a known recording fraction $\rho$, complete-path target vectors and optional coarsening fibres. Recorded and missing masses satisfy a marginal recording-odds restriction with sensitivity parameter $\kappa\geq1$. A conditional target has the form $c^\top q/d^\top q$. The Charnes--Cooper transformation sets $t=(d^\top q)^{-1}$ and $v=tq$, so that $d^\top v=1$ and the objective becomes $c^\top v$ \citep{Charnes1962}. The solver checks primal feasibility and dual agreement and returns endpoint-attaining laws. Optional recorded counts produce simultaneous binomial confidence constraints and an outer confidence projection. EIF scores do not replace these recorded-law constraints. Positive conditioning denominators, finite declared support and the known recording fraction are essential.

For large supports, \code{cs\_bounds\_sparse()} avoids the dense Charnes--Cooper tableau by exploiting the existing recording-odds constraints. Within a recording fibre, the linear objective can concentrate on its maximizing atom. A two-resource sorted allocation solves the resulting recorded/unrecorded mass problem. Parametric fractional iterations \citep{Dinkelbach1967} supply feasible endpoint laws and reconstructible dual certificates. The outward endpoint is checked against an attained value. Work per iteration is $O(J+M\log M)$ and memory $O(J+M)$ for $J$ complete atoms and $M$ recorded categories. This exact structural reduction permits arbitrary partition fibres and the implemented recorded-mass confidence boxes; it is not a generic solver for additional arbitrary cross-fibre restrictions. A conditioning denominator bounded away from zero is required for the fractional certificate. The explicitly labelled support-envelope fallback remains for a zero denominator lower bound.

\subsection{Future covariate evolution}
Future evolution is selected by \code{covariate\_evolution}. The default \code{"freeze"} uses landmark values throughout the horizon. The option \code{"locf"} without a future schedule agrees with freeze; with \code{newdata\_future}, each declared update is applied prospectively and held until the next update. Such schedules are forecast scenarios, not retrospectively imported future measurements. Generator segments split at their update times, and one frailty draw is retained across the complete ordered propagation.

For \code{"joint\_submodel"}, \code{cs\_covariate\_model()} estimates Gaussian linear random-coefficient models by ML or REML, following the longitudinal framework of \citet{Laird1982}. Prediction requires \code{covariate\_history} with actual measurement times through the landmark. Repeated event-interval rows and latent quadrature times are not treated as additional biomarker measurements. For one covariate the conditional random-effect distribution is
\begin{equation}\label{eq:covariate}
 \begin{aligned}
 W_i&=X_i\beta+Z_i b_i+e_i,\quad b_i\sim N(0,D),\quad e_i\sim N(0,\sigma^2 I),\\
 b_i\mid W_i&\sim N\!\left(DZ_i^\top V_i^{-1}(W_i-X_i\beta),
 D-DZ_i^\top V_i^{-1}Z_iD\right),\quad V_i=Z_iDZ_i^\top+\sigma^2I.
 \end{aligned}
\end{equation}
Here $W_i$ is the past measurement vector for subject $i$, $X_i$ and $Z_i$ are fixed- and random-effect design matrices, $\beta$ the fixed coefficient vector, $b_i$ the random intercept and optional slope, $D$ its covariance, $e_i$ measurement noise, $\sigma^2$ its variance, $I$ the identity and $V_i$ the marginal measurement covariance. These $X_i,Z_i$ are local mixed-model matrices, not the frailty or intensity notation in Section~\ref{sec:model}. Gaussian conditioning gives the displayed distribution. Drawn future latent covariate curves are propagated and averaged; the reported Monte Carlo error is conditional on estimated submodel parameters. Independent covariate submodels, no feedback from future events and correctly specified evolution remain assumptions. This is not a jointly estimated measurement-error survival model, and it does not automatically include longitudinal parameter uncertainty.

\subsection{Alternative frailty laws and diagnostics}
The choice \code{frailty = "gamma"}, \code{"lognormal"} or \code{"discrete"} changes the mixing distribution while retaining a mean-one scale. Gamma integration remains analytic. Lognormal integration uses Gauss--Hermite nodes with $\log Z\sim N(-v/2,v)$ and $v=\log(1+\phi)$, where $Z$ is frailty and $\phi$ its variance. The discrete option estimates two to eight mass points and their weights. With unconstrained weight parameters $a_\ell$ and log-locations $v_\ell$, it uses
\begin{equation}\label{eq:discrete}
 w_\ell=\frac{e^{a_\ell}}{\sum_{m=1}^L e^{a_m}},\qquad
 z_\ell=\frac{e^{v_\ell}}{\sum_{m=1}^L w_m e^{v_m}},\qquad
 \Prb(Z=z_\ell)=w_\ell.
\end{equation}
Here $L$ is the chosen number of points, $\ell,m$ index those points, $w_\ell$ is a probability, $z_\ell$ a positive frailty value and $Z$ the random frailty. The construction imposes $\sum_\ell w_\ell z_\ell=1$. Estimated mass-point mixing follows the motivation of \citet{Heckman1984}, but a fixed finite number of points is not an unrestricted nonparametric distribution estimator. Derivatives of weights and normalized locations enter the likelihood score and observed information; Louis's correction integrates both path and frailty uncertainty. Near-coincident points or vanishing weights can yield nonregular fits and are not declared regular merely because the optimizer stops. Lognormal fitting doubles quadrature nodes with warm-started optimization and checks changes in individual log likelihoods and mean scores. Reaching the declared maximum without convergence marks integration unresolved. This is a numerical refinement diagnostic, not a rigorous quadrature error bound.

\code{cs\_frailty\_diagnose()} fits the zero-variance null and forms a nuisance-adjusted one-sided variance score. Its half-point-mass/half-$\chi^2_1$ reference is used only under a correctly specified regular null with an identifiable scalar variance direction, consistent with the boundary principles of \citet{Self1987}. The null nuisance fit itself must converge with full-rank positive information, small mean score and no numerical coefficient-bound contact; otherwise the diagnostic retains its statistic but reports no $p$ value. Nonunit case weights use a derivative-based nuisance correction and require fixed or exogenous weights and a valid weighted-score central limit theorem. Alternative fits retain the same weights. Gamma, lognormal and estimated discrete alternatives are nonnested or singular comparisons. The interface reports their likelihoods, parameter counts, AIC and regularity flags; it does not assign an ordinary chi-square likelihood-ratio reference to those comparisons. AIC is descriptive, particularly for nonregular mixtures; weighted objective values need not have ordinary AIC's risk interpretation. The verification supplement gives both weighting branches.

\subsection{Calibration and sparse-risk uncertainty}
\code{cs\_regularize()} adds one-half the log determinant of observed information to an eligible exact-history gamma likelihood. The update stays inside a positive-information domain, uses an analytic trace score and estimates penalized curvature by central differences. It is a local regularizer; automatic first-order Firth bias cancellation for this nonlinear frailty family is not asserted. \code{cs\_band()} draws Gaussian parameter perturbations using a supplied sandwich covariance, evaluates the nonlinear forecast and forms simultaneous maximum-$t$ bands on the logit or identity scale. It is a frequentist approximation, not a literal posterior. Invalid draws cause a visible failure and are not clipped or replaced.

Weighted conformal inversion is supplied by \code{cs\_conformal()}. It retains the candidate's weight and handles score ties, following the weighted-exchangeability formulation of \citet{Tibshirani2019}. Known weights, estimated weights and externally certified weight bands have different output labels. Estimated weights alone do not confer exact coverage. The routine inverts a supplied candidate grid; it is not a general continuous-outcome set-search engine. \code{cs\_pivot()} instead inverts a model-specific exact pivot for independent uncensored terminal times with known dispersion and constant baseline. \code{cs\_cornish\_fisher()} implements an explicitly asymptotic quantile expansion. Their assumptions are intentionally not hidden behind a single generic confidence argument.

\section[Using coherentSurv]{Using \pkg{coherentSurv}}\label{sec:using}
The source distribution installs with a C++17 toolchain and \pkg{Rcpp}. Suggested packages support public-data examples, plots and vignettes. The three synthetic datasets are small and distributed with the package. The following complete example uses monthly time and an explicit assessment origin:
\begin{CodeChunk}
\begin{CodeInput}
R> library("coherentSurv")
R> data("sim_perinatal")
R> fit <- cs_fit(~ x, sim_perinatal, event_types = c("1", "2"),
+                  time_unit = "months")
R> glance(fit)
R> one <- sim_perinatal[sim_perinatal$id == 1, ]
R> pred <- predict(fit, one, landmark = 0,
+                 horizon = c(0.25, 0.5, 1), type = "all")
R> pred[, c("unit", "horizon", "target", "estimate")]
\end{CodeInput}
\end{CodeChunk}
Table~\ref{tab:quickstart} is generated by this workflow. The three numerical domains use the same data contract, with different declared units and generating rates. Domain names are illustrative and do not imply that the simulated covariate is a clinically measured exposure or a validated engineering failure mode.
\begin{table}[tbp]
\centering\small
\begin{tabular}{rrrrrr}
\toprule
Horizon & Terminal risk & Survival & Burden 1 & Burden 2 & Time alive \\
\midrule
0.25 & 0.0547 & 0.9453 & 0.1478 & 0.0729 & 0.2431 \\
0.50 & 0.1062 & 0.8938 & 0.2917 & 0.1436 & 0.4729 \\
1.00 & 0.2000 & 0.8000 & 0.5660 & 0.2774 & 0.8958 \\
\bottomrule
\end{tabular}
\caption{Forecasts for the first synthetic perinatal subject at the assessment origin. Burdens are expected counts; restricted survival is in months.}\label{tab:quickstart}
\end{table}

The fitted object supports standard S3 methods, including \code{print()}, \code{summary()}, \code{coef()}, \code{vcov()} and \code{logLik()}. The additional \code{tidy()}, \code{glance()} and \code{augment()} interfaces organize coefficients, model summaries and forecasts. Predictions include target names, units and numerical diagnostics; plots preserve these labels. Audit metadata record the call, omitted units, data fingerprint and assumptions. The fingerprint is an integrity check, not a cryptographic privacy guarantee.

The 0.10.1 wrappers expose the no-frailty, exact-observation special case without changing the numerical core. \code{cs\_surv()} accepts either a one-sided predictor formula with named time/status columns or a supported \code{Surv(time, status)} response. The binary status is zero for censoring and one for the event. The model is piecewise-exponential proportional hazards, using the full likelihood; it is not Cox partial-likelihood estimation. The explicit \code{frailty="none"} alias fixes frailty variance at zero in the main interface. Start--stop data require a unit identifier and contiguous observation from the origin; delayed entry remains unsupported by these wrappers.

\code{cs\_competing()} fits no-frailty cause-specific hazards for declared mutually exclusive absorbing causes. It combines every fitted cause hazard in one sparse generator, returning \code{cif\_<cause>}, \code{survival}, all-cause \code{risk} and \code{rmst}. Thus cumulative incidence is not obtained by treating each cause's isolated survival complement as its probability. A declared cause with no observed events fails visibly instead of receiving a regular fitted interval. The joint sandwich covariance uses aligned subject scores; conditional CIF confidence bands are not generated automatically. Competing forecasts support frozen covariates or explicit future schedules, not a latent longitudinal submodel.

For data frames \code{d} and \code{dc} containing the declared columns, the calls are:
\begin{CodeChunk}
\begin{CodeInput}
R> fit_s <- cs_surv(~ x, d, time = "time", status = "status",
+                  unit = "id", time_unit = "years")
R> fit_c <- cs_competing(~ x, dc, time = "time", status = "cause",
+                       unit = "id", causes = c("1", "2"),
+                       censor_code = 0, time_unit = "years")
R> predict(fit_c, landmark = 0, horizon = 1, type = "all")
\end{CodeInput}
\end{CodeChunk}
Here \code{dc\$cause} uses zero for censoring and the declared labels for absorption. Forecast tables add the \code{cs\_forecast} class while retaining \code{coherentSurv\_prediction}; forecasts are requested through the existing \code{predict()} generic. Covariate-only prediction profiles are permitted when the analyst explicitly supplies eligible event-free profiles at the landmark; their temporary mapping intervals are not observed follow-up.

The competing-object \code{validate()} method evaluates one declared cause's cumulative incidence. Earlier competing events are observed noncases, with censoring weights evaluated at their event times; they are not recoded as loss to follow-up. The default still requires marginal independent censoring. Externally supplied conditional censoring probabilities need the documented event-time and horizon alignment. Metrics remain held-out point estimates rather than automatic CIF confidence bands.

The executable \code{replication/standard\_regimes.R} checks a two-cause example against closed-form probabilities and applies the single-event wrapper to \pkg{survival}'s lung data. The targeted addendum reports the numerical agreement, held-out metrics and exact data-coding choices; these examples do not add a claim of clinical or causal validation.

\begin{figure}[t]
\centering\includegraphics[width=.85\textwidth]{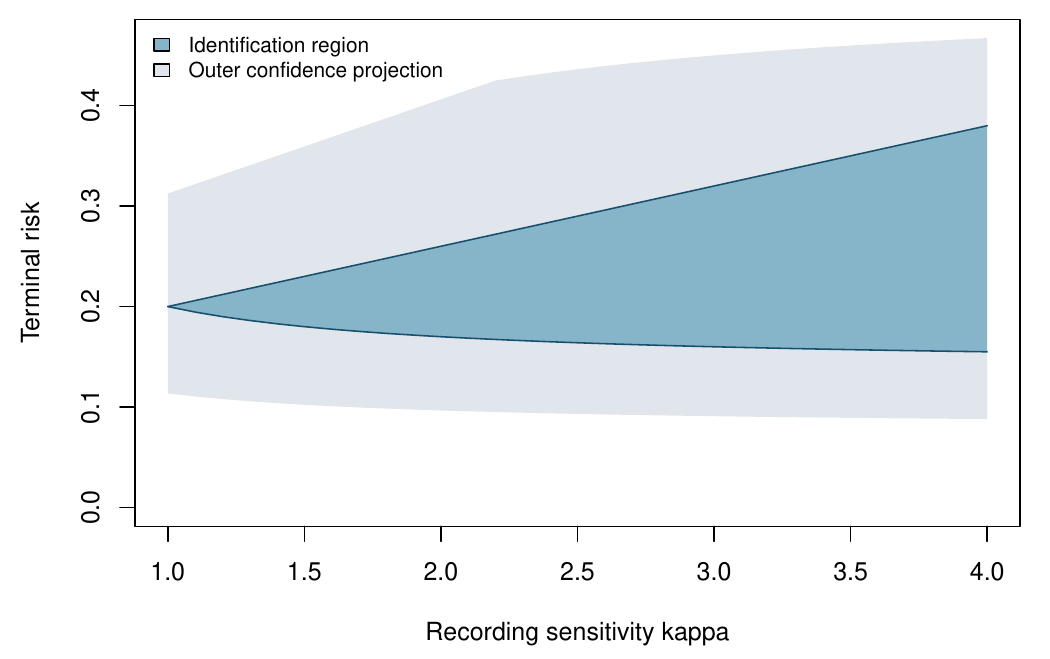}
\caption{Identification bounds and simultaneous outer confidence limits in a three-atom example. The known recording fraction is 0.7. Wider sensitivity classes admit additional complete laws; sampling uncertainty is projected separately.}\label{fig:bounds}
\end{figure}

A finite-support sensitivity calculation has a different input contract:
\begin{CodeChunk}
\begin{CodeInput}
R> count <- c(50, 30, 20)
R> sensitivity <- cs_bounds(count / sum(count), c(0, 0, 1),
+     kappa = 2, recording_fraction = 0.7, counts = count)
R> c(sensitivity$lower, sensitivity$upper)
R> c(sensitivity$outer$lower, sensitivity$outer$upper)
\end{CodeInput}
\end{CodeChunk}
The first interval is an identification region at the empirical recorded law; the second adds sampling uncertainty. Figure~\ref{fig:bounds} varies the sensitivity parameter without changing the observed counts. Its widths describe this declared experiment and should not be interpreted as a biological decomposition of uncertainty.

\subsection{From event records to a field-specific report}
The reusable object is an event history, not a disease or industry label. The analyst defines the independent unit, time origin, recurrent marks, absorbing endpoint, observation process and forecast horizon before fitting. \code{predict(..., type = "all")} then returns terminal risk, survival, expected recurrent counts and restricted time alive from one fitted law. Expected counts are not occurrence probabilities. Table~\ref{tab:reporting} connects those outputs to a report; the companion cross-field guide gives formulas, exact field names, inputs and examples.

\begin{table}[tbp]\centering\small
\begin{tabular}{>{\raggedright\arraybackslash}p{.22\textwidth}>{\raggedright\arraybackslash}p{.35\textwidth}>{\raggedright\arraybackslash}p{.33\textwidth}}\toprule
Question & Native output or interface & Minimum reporting context \\ \midrule
Future endpoint and burden & \code{predict()}: \code{risk}, \code{survival}, \code{burden\_*}, \code{rmst}; \code{estimate} & Unit, landmark, horizon, event eligibility, time unit and covariate-evolution rule. \\
Population inference & Targeted \code{predict()}: \code{std\_error}, \code{lower}, \code{upper}, \code{interval}, \code{inference\_status} & Population target versus individual forecast; regularity, nuisance, censoring and selection qualifications. \\
Recording sensitivity & \code{cs\_bounds()} or \code{cs\_bounds\_sparse()}: endpoint bounds and optional outer projection & Declared support/fibres, recording fraction, sensitivity parameter, denominator and category counts. \\
Prediction error and ranking & \code{validate()}: \code{Brier}, \code{AUROC} & Independent evaluation units, fixed horizon, observed events and censoring weights. \\
Calibration & \code{CITL}, \code{calibration\_slope}, \code{ICI} & Case/control counts, calibration-fit status and local smoothing convention. \\
Threshold decisions & \code{sensitivity}, \code{specificity}, \code{PPV}, \code{NPV}, \code{net\_benefit} & Prespecified threshold, weighted confusion table, censoring model and relevant decision consequences. \\
Numerical reliability & Fit convergence/information; series, mixture-compression and covariate-MC diagnostics & Numerical errors are reported separately from confidence intervals and prediction error. \\
External reporting calculations & Companion reporting functions or analyst code & Cost/utility coefficients, precision--recall convention, bootstrap scope or simulation oracle, as applicable. \\
\bottomrule\end{tabular}
\caption{A reporting contract for the current package. Native validation has ten metrics. Additional reporting calculations do not become package exports or empirical benefits merely by appearing in a report.}\label{tab:reporting}
\end{table}

The companion replication archive also contains \code{replication/research\_metrics.R}, which must be sourced explicitly. Its \code{cs\_research\_report()} adds censoring-weighted average precision, threshold curves, comparator net benefits and optional fixed misclassification losses; \code{cs\_value\_burden()} translates expected counts into an additive valuation using supplied costs. These functions are companion analysis code, not package exports. The optional independent-validation-unit bootstrap holds the trained predictor fixed, retains undefined draws and withholds intervals when its stated reporting requirements fail. It does not provide simultaneous or exact confidence coverage. Valuation reports no joint uncertainty unless the analyst supplies and propagates appropriate joint forecast draws.

In hospital research, a person can have recurrent admissions or clinically prespecified event marks before death; the report gives expected event counts, time alive and estimated horizon risk, with calibration assessed separately. In trials and policy research, the same outputs can support prognostic stratification, follow-up planning or evaluation of a prespecified prediction rule; randomized origin alone does not turn these predictions into treatment-effect estimates. In economics or banking, an eligible account may have repeated delinquency events before a prespecified absorbing default, while operating assets may experience recurrent faults before irreparable failure in engineering. These are mappings of the data contract, not external bank or industrial validation. Costs, loss given default, quality-of-life weights and intervention effects require additional inputs and an appropriate study design.

Validation distinguishes probability accuracy, ranking and decision consequences. The censoring-weighted Brier score follows survival prediction-error principles \citep{Graf1999}; horizon AUROC concerns event cases by that horizon versus subjects still event-free \citep{Heagerty2000}. The implemented ICI is a weighted absolute difference from a local calibration curve \citep{Austin2019}; net benefit is a threshold-weighted true-positive/false-positive contrast \citep{Vickers2006}. None alone demonstrates causal benefit. Native \code{validate()} returns point estimates and defined/status flags, not confidence intervals. Validation-set resampling can quantify uncertainty conditional on fitted predictions; model-development uncertainty requires repeating the complete training procedure. Whole-subject splits must prevent history leakage, and evaluation must be repeated in the intended setting before operational use.

\begin{table}[t]
\centering\small
\begin{tabular}{p{0.18\textwidth}p{0.22\textwidth}p{0.22\textwidth}p{0.22\textwidth}}
\toprule
Contract & Clinical study & Fleet reliability & Credit study \\
\midrule
Independent unit & Participant or patient & Device & Account or borrower \\
Recurrent mark & Defined admission or toxicity episode & Fault or repair episode & Delinquency or covenant episode \\
Absorbing endpoint & Death & Irreversible failure & Default, if absorbing by definition \\
Terminal risk & Death probability & Failure probability & Default probability \\
Expected burden & Episodes before death & Episodes before failure & Episodes before default \\
Restricted time alive & Time alive within the horizon & Time before failure within the horizon & Time before default within the horizon \\
\bottomrule
\end{tabular}
\caption{Examples of a common event-history contract. These are domain translations, not three externally validated applications. Time units, observation rules and the independent unit must be declared for each study.}
\label{tab:domain_contract}
\end{table}

Table~\ref{tab:domain_contract} translates the core targets without changing their probability definitions. The additional executable tutorials provide synthetic fleet, credit and randomized-trial examples using the installed API. A separate guide checks label-invariance of the trajectory representation. These examples preserve the distinction between timing coarsening with known counts and incomplete recording of unknown episodes. Restricted time alive is not automatically event-free time, quality-adjusted survival or unrestricted mean failure time. Expected distress counts are not financial expected loss without an exposure, loss-given-default or other explicit reward model. Terminal absorption is an analysis definition, distinct from administrative loss to follow-up and from a trial's intercurrent-event strategy.

\section{Computational verification and benchmarks}\label{sec:benchmarks}
Verification has three levels. Algebraic unit tests compare derivatives with independent numerical differences and propagation with closed forms. Cross-language fixtures compare the new routines with unchanged numerical outputs from the Version 5 research implementation. Workflow tests check subject separation, observation restrictions, probability conservation, future-information exclusion and undefined metric handling. Table~\ref{tab:parity} reports actual discrepancies; small numerical differences verify implementation agreement, not statistical coverage under arbitrary laws.
\begin{table}[tbp]
\centering\small
\begin{tabular}{lr}
\toprule
Comparison & Maximum absolute difference \\
\midrule
Likelihood score vs numerical derivative & 2.37e-12 \\
Observed information vs score derivative & 3.45e-13 \\
Louis information vs score derivative & 2.81e-13 \\
Probability vs analytic propagation & 7.77e-16 \\
Reward vs analytic integral & 4.22e-15 \\
Ordered semigroup agreement & 1.67e-16 \\
Finite-path target vs Python oracle & 4.44e-16 \\
Individual EIF vs Python oracle & 2.72e-15 \\
Conditional LP endpoint vs Python oracle & 1.11e-16 \\
Outer projection vs Python oracle & 3.89e-15 \\
Observed-information penalty score vs oracle & 6.68e-15 \\
\bottomrule
\end{tabular}
\caption{Executed numerical checks. Each row concerns its stated fixture and comparison, not a universal accuracy or coverage guarantee.}\label{tab:parity}
\end{table}

The cross-fitted fixture has 800 subjects, 13 legal paths and four target functions. The LP fixtures include conditional targets, coarsening fibres and confidence projection. They are deterministic verification examples, not new Monte Carlo coverage studies. Earlier large simulation results in the methodological archive are retained as reference evidence and are not relabelled as new package runs. Tables~\ref{tab:parity}--\ref{tab:performance} and the five case studies preserve the measured 0.9.1 predecessor results; new 0.10.0 verification is reported separately below. The companion script records source hashes before performance measurement.

\begin{figure}[t]
\centering\includegraphics[width=.85\textwidth]{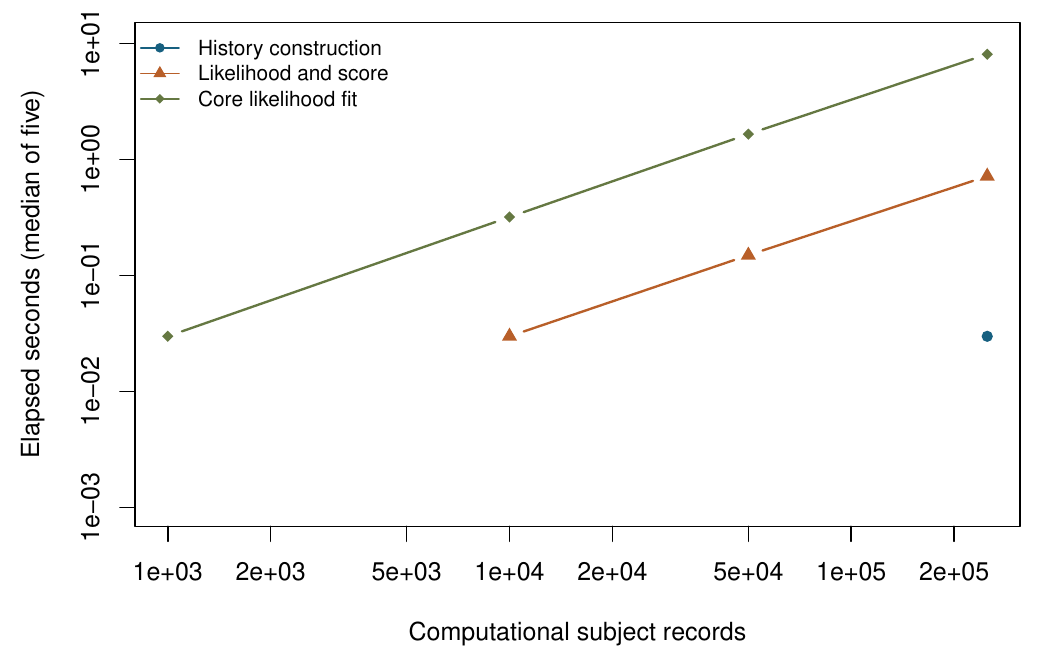}
\caption{Measured elapsed time under the frozen benchmark protocol. Lines connect five-run medians for the specified workload; they are not complexity proofs or comparisons of unmatched statistical models.}\label{fig:runtime}
\end{figure}

The performance protocol fixes four sample sizes and five repetitions per size. It times history construction, likelihood-and-score evaluation, and low-level optimization separately. Larger workloads repeat a fixed synthetic history template under distinct unit identifiers; these are computational workloads, not independent Monte Carlo cohorts. The benchmark model has a fixed small coefficient dimension, so its timings should not be extrapolated to a growing-dimensional history design. A separate high-level fit includes R data preparation. The measured machine is an AMD Ryzen 9 3900X with 12 physical cores and 24 logical processors, running Windows 11; the timed numerical workload uses one thread. Table~\ref{tab:performance} reports medians and Figure~\ref{fig:runtime} displays scaling. Process memory includes preparation and is measured separately from model object size.
\begin{table}[tbp]
\centering\small
\begin{tabular}{rrrrr}
\toprule
Subjects & History (s) & Score (s) & Core fit (s) & Peak GB \\
\midrule
1,000 & <0.01 & <0.01 & 0.030 & 0.171 \\
10,000 & <0.01 & 0.030 & 0.320 & 0.156 \\
50,000 & <0.01 & 0.150 & 1.660 & 0.234 \\
250,000 & 0.030 & 0.720 & 8.110 & 0.771 \\
\bottomrule
\end{tabular}
\caption{Median elapsed seconds from five repeated runs at each sample size. History preparation, a likelihood-and-score evaluation and a four-parameter core fit are distinct operations. Memory is peak working set of a fresh process including preparation.}\label{tab:performance}
\begin{minipage}{\textwidth}\footnotesize An elapsed value below timer resolution is reported as a bound. These are workload measurements on the stated host; they are not timings for all high-level data transformations or arbitrary model sizes.\end{minipage}
\end{table}

The protocol's two-second and 45-second objectives are evaluated against the matching measured operation, not against a conveniently shorter kernel. The same model's independent R implementation provides a numerical reference. Companion-package fits are reported only as executable examples because their statistical specifications differ. Five-, 15- and 30-node gamma quadrature calculations quantify sensitivity to integration resolution; agreement at one profile does not certify a uniform quadrature bound.
The executed high-level fit at 1000 subjects had a median elapsed time of 1.66 seconds. Larger high-level fits were not timed under this protocol, so the 45-second objective for a full 250,000-subject workflow remains unverified. The small core fit and process-memory measurements must not be substituted for that missing comparison.
Five-node gamma quadrature differed from the 30-node reference by at most $1.48\times10^{-5}$ across the tested coordinates, exceeding the registered $10^{-5}$ tolerance. Fifteen nodes gave a maximum difference of $7.93\times10^{-10}$. The default remains 30 nodes; this profile-specific comparison does not certify a uniform integration bound.

\subsection{Verification of the five extensions}
Table~\ref{tab:revision} distinguishes implementation agreement, numerical integration diagnostics and empirical inferential validation. The automatic fibre generator retained all 929 colon participants and five same-day pairs, producing the same 7,447 paths as the specialized predecessor adapter. Two overlapping same-mark windows were also checked against a direct union integral to detect permutation double counting. In a deliberately large-order fibre, 200 independent importance integrations with 2,000 attempted draws each had empirical standard deviation 0.1513 and mean reported MCSE 0.1502. This diagnostic correctly reveals material integration noise at that budget; it is not confidence coverage for a statistical estimand.

The large-support solver agreed with 80 independent SciPy/HiGHS Charnes--Cooper problems. Three of these were feasible boxes that the old dense solver had classified as infeasible; the new reduction and independent oracle agreed. Sparse scale checks reached 50,000 atoms and 50,000 states. These deterministic instances establish execution at the declared size, not universal speed across graphs. On this machine, the native 50,000-state calculation used an estimated 4.40 MB workspace and 15 Poisson terms. The end-to-end R bound-plus-projection call at 50,000 atoms took 0.06 seconds in the recorded run; compilation and unrelated data preparation are excluded.

\begin{table}[tbp]\centering\small
\begin{tabular}{>{\raggedright\arraybackslash}p{.23\textwidth}>{\raggedright\arraybackslash}p{.48\textwidth}>{\raggedright\arraybackslash}p{.20\textwidth}}\toprule
Module & Recorded comparison or diagnostic & Result \\ \midrule
Automatic fibres & 929 colon subjects; log likelihood vs specialized adapter & $3.14\times10^{-13}$ \\
Louis information & Generated paths vs specialized adapter & $1.14\times10^{-13}$ \\
Sparse bounds & 80 independent HiGHS problems; maximum endpoint difference & $1.72\times10^{-15}$ \\
Large-support bounds & 50,000 atoms; fractional certificate gap & $4.28\times10^{-18}$ \\
CSR propagation & 50,000 states; probability series bound & $3.74\times10^{-14}$ \\
Continuous targeting & Individual EIF vs unchanged V5 oracle & $7.77\times10^{-16}$ \\
Future schedule & Ordered forecast vs analytic oracle & $1.75\times10^{-15}$ \\
Lognormal frailty & 40-node log likelihood vs adaptive integration & $1.98\times10^{-13}$ \\
\bottomrule\end{tabular}
\caption{New 0.10.0 verification. Differences measure numerical agreement; series bounds concern real-arithmetic truncation. They are distinct from empirical confidence coverage.}\label{tab:revision}
\end{table}

The scheduled-covariate forecast agreed with a closed-form ordered-intensity calculation to $1.75\times10^{-15}$. In a Gaussian future-covariate example with 500 common random-effect draws, terminal-risk estimates at grids of 8, 16 and 32 intervals were 0.17273, 0.17339 and 0.17373. Their conditional Monte Carlo standard errors were 0.00088, 0.00089 and 0.00090. Grid refinement and Monte Carlo integration are separate error sources; neither is a confidence interval for the fitted model. At 40 lognormal quadrature nodes, a fixed-parameter log likelihood differed from adaptive numerical integration by $1.98\times10^{-13}$.

The separately registered targeting experiment generated 5,000 cohorts in each of four cells: sample sizes 250 and 1,000 crossed with ordinary and sparse event rates. Histories were generated by a history-dependent finite-state law without latent frailty; the fitted event nuisance was pooled and censoring nuisance estimated by baseline stratum. The four simultaneous targets were two recurrent burdens, terminal risk and restricted survival. All 20,000 generated cohorts remain in the denominators, including unsuccessful score or positive-variance checks. No execution exception occurred.

\begin{table}[tbp]\centering\small
\begin{tabular}{lrrrr}\toprule
Regime & Subjects & Raw Wald & Transformed Gaussian & Numerical success \\ \midrule
Ordinary & 250 & 0.9314 (0.0036) & 0.9510 (0.0031) & 5000/5000 \\
Ordinary & 1000 & 0.9460 (0.0032) & 0.9516 (0.0030) & 5000/5000 \\
Sparse & 250 & 0.5582 (0.0070) & 0.5942 (0.0069) & 4773/5000 \\
Sparse & 1000 & 0.8612 (0.0049) & 0.9254 (0.0037) & 4972/5000 \\
\bottomrule\end{tabular}
\caption{Simultaneous coverage in the registered 0.10.0 validation. Parentheses give binomial Monte Carlo standard errors. Raw comparisons precede the public zero-event regularity guard; all 5,000 cohorts per cell are retained.}\label{tab:continuous_validation}
\end{table}

\begin{figure}[tbp]\centering\includegraphics[width=\textwidth]{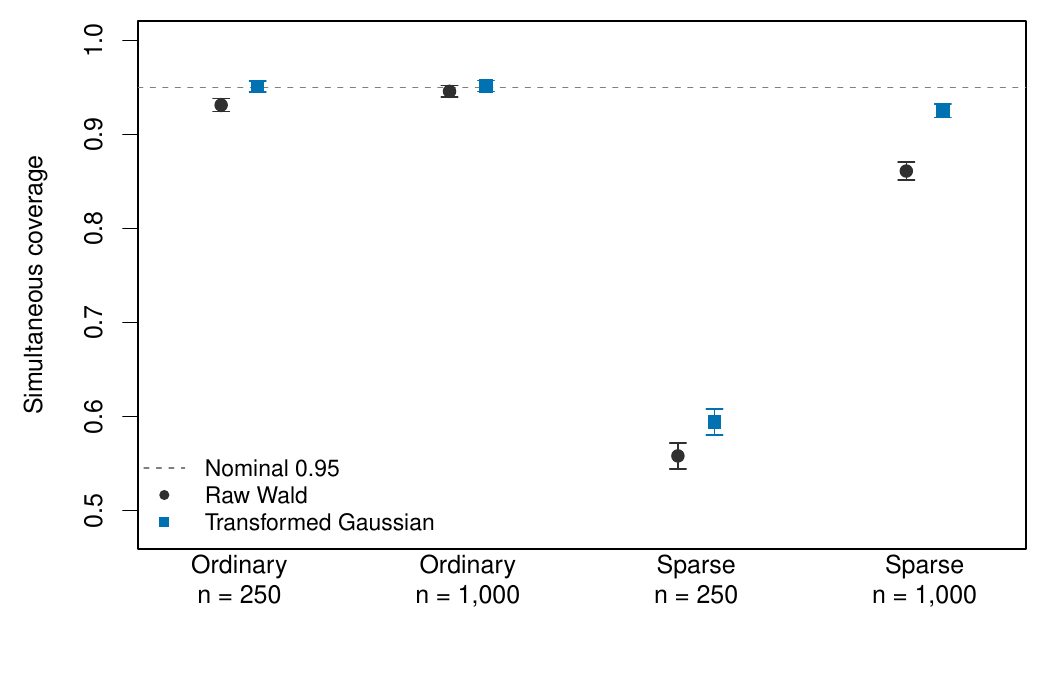}
\caption{Registered simultaneous-coverage comparison. Error bars are 1.96 Monte Carlo standard errors; the dashed line is the 0.95 nominal level. Sparse failures remain visible.}\label{fig:continuous_validation}\end{figure}

Transformed intervals approached nominal coverage in the ordinary cells but did not repair the sparse-event failures (Table~\ref{tab:continuous_validation} and Figure~\ref{fig:continuous_validation}). At 250 sparse subjects, simultaneous coverage was 0.5942 rather than 0.95; at 1,000 it was 0.9254. The simulation therefore supports a qualified regular-regime interface, not universal sparse-cell validity. The public API now withholds regular Gaussian limits when a target has an observed zero-event boundary. This diagnostic is additional to the preserved raw comparison and is not retrospectively scored as improved coverage.

The public fallback uses observed lower and upper reward envelopes under censoring. For a target with known support, Hoeffding inequalities and a union bound give a simultaneous conservative outer confidence set; this argument requires independent identically distributed units but does not require independent censoring. Terminal-event indicators and restricted survival have such bounds. Uncapped burdens receive the uninformative outer support $[0,\infty)$, rather than a fabricated finite upper limit. The output is labelled \code{bounded\_coarsening\_projection}; it is distinct from the sharp finite-support sensitivity solver and does not recover point identification. Explicit Gaussian requests at a zero-event boundary remain unavailable. For the stated finite-sample guarantee, request \code{interval = "projection"} uniformly. Data-dependent automatic switching between approximate Gaussian and projection intervals has neither the unconditional guarantee of the projection-only procedure nor a selection-conditional guarantee from this argument. Numerical integration tolerances, MCSE and confidence coverage remain separate diagnostics.
A separately registered envelope experiment generated 5,000 cohorts under each of complete observation, independent censoring and informative censoring. Joint coverage of the bounded coordinates was 0.9992, 1.0000 and 1.0000, respectively. The first estimate has MCSE 0.00040; the two all-covered cells have binomial Monte Carlo lower limits 0.99926, rather than evidence of literal certainty. Median risk widths were 0.2960, 0.6380 and 0.6480, and restricted-survival widths were 0.2406, 0.4291 and 0.4329. These conservative sets trade precision for weaker observation assumptions. The 15,000-cohort check does not replace or retrospectively repair the separate 20,000-cohort Gaussian experiment.

\Needspace{12\baselineskip}
\section{Reproducible case studies}\label{sec:cases}
\begin{figure}[t]
\centering\includegraphics[width=\textwidth]{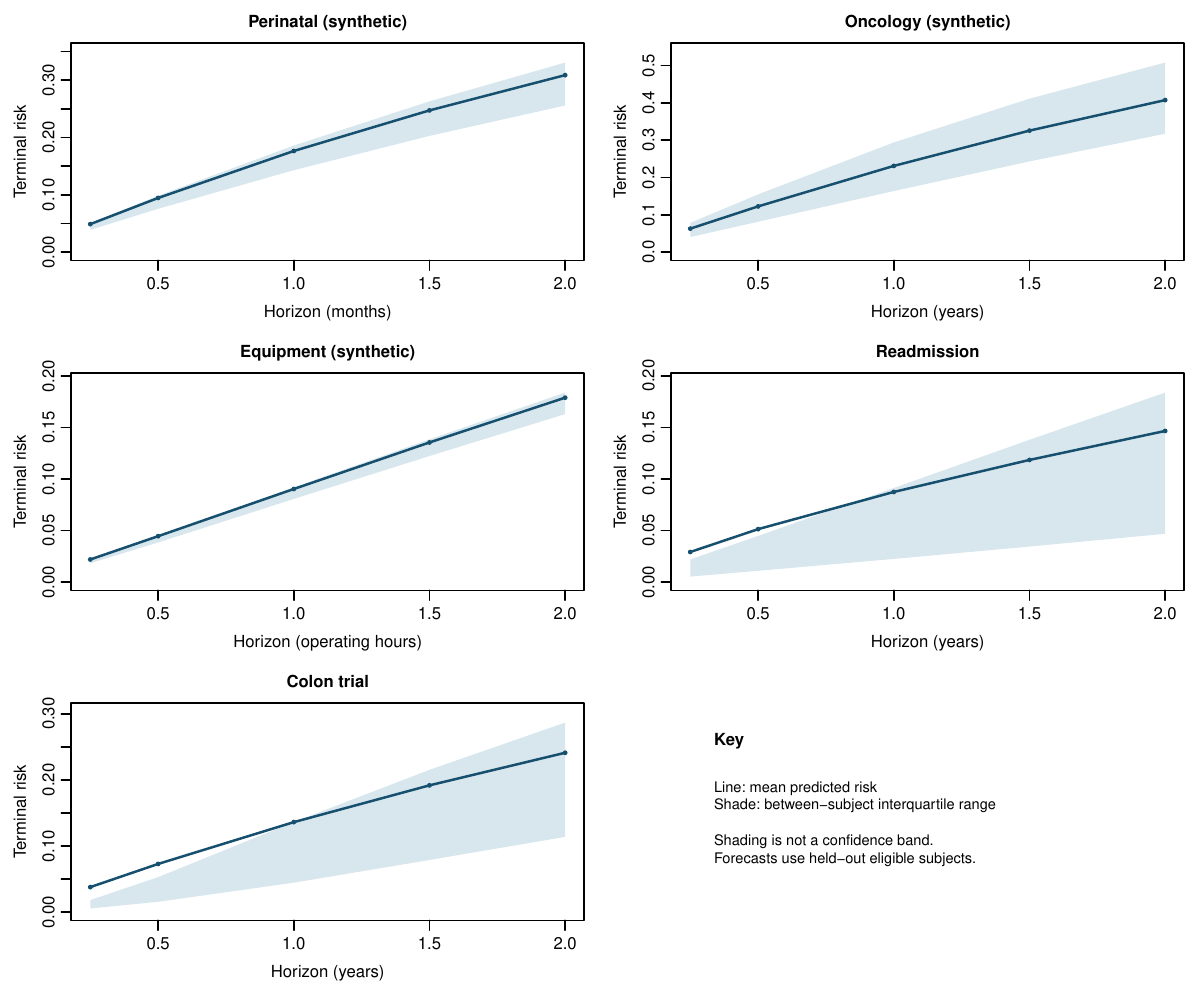}
\caption{Common-law forecasts in the five case studies. Curves summarize the generated examples under their stated history and observation specifications. They are prognostic model outputs, not intervention effects. Synthetic cases are explicitly labelled.}\label{fig:cases}
\end{figure}

Each case records data provenance, an estimand, a forecast summary, held-out validation and a limitation. Table~\ref{tab:cases} summarizes the fitted samples and diagnostics. The real-data adapters obtain records from their canonical source packages. The software source distribution contains only synthetic records and test fixtures; aggregate case results are sufficient for the manuscript. Figure~\ref{fig:cases} shows the generated forecast summaries, with time units and eligibility carried from each analysis.
\begin{table}[tbp]
\centering\small
\begin{tabular}{lrrrrl}
\toprule
Case & Subjects & Recurrences & Terminal & Frailty variance & Status \\
\midrule
Perinatal (synthetic) & 160 & 335 & 59 & 0.485 & converged \\
Oncology (synthetic) & 160 & 248 & 68 & 0.398 & converged \\
Equipment (synthetic) & 160 & 398 & 41 & 0.618 & converged \\
Readmission & 403 & 458 & 109 & 0.751 & converged \\
Colon trial & 929 & 468 & 452 & 0.693 & converged \\
\bottomrule
\end{tabular}
\caption{Full-data case fits. Event counts are source events, not the number of interval rows or quadrature paths. All figures and validation statistics retain the declared observation model.}\label{tab:cases}
\end{table}

\begin{table}[tbp]
\centering\small
\begin{tabular}{llrrrr}
\toprule
Case & Time unit & Risk & Burden 1 & Burden 2 & Time alive \\
\midrule
Perinatal (synthetic) & months & 0.177 & 0.610 & 0.261 & 0.907 \\
Oncology (synthetic) & years & 0.231 & 0.525 & 0.208 & 0.880 \\
Equipment (synthetic) & operating hours & 0.090 & 0.615 & 0.253 & 0.955 \\
Readmission & years & 0.087 & 0.316 & -- & 0.951 \\
Colon trial & years & 0.136 & 0.143 & -- & 0.929 \\
\bottomrule
\end{tabular}
\caption{Mean held-out forecasts at a duration of one stated time unit. Burden is an expected count; time alive is measured in the indicated unit. A dash denotes an event type absent from that case, not an estimated zero.}\label{tab:caseforecasts}
\end{table}

\subsection{Synthetic perinatal histories}
The perinatal example contains two recurrent marks and one absorbing endpoint on a monthly scale. Its generating law has shared gamma frailty and capped-count feedback. The applied interpretation is expected additional morbidity episodes and terminal risk over a defined follow-up window, while the data themselves remain simulated rather than maternal records. The forecast table reports all target types from one fitted law. Held-out Brier score and discrimination demonstrate the validation interface; their values in Table~\ref{tab:validation} concern this generating mechanism only. The marks are deliberately generic and do not establish a clinical sequence between distinct diseases.

\subsection{Synthetic oncology histories}
The oncology demonstration uses years and a different event-rate configuration. It distinguishes recurrent episodes from terminal risk and restricted time alive. The same schema and function calls apply without renaming the mathematical targets. Validation is performed on held-out subjects rather than repeated records from the training subjects. This example demonstrates transfer of the interface to a different time scale and intensity configuration; it does not establish transportability to a cancer population or identify a treatment effect.

\subsection{Synthetic equipment histories}
The robotics demonstration uses operating hours and recurrent repair-type marks before terminal failure. Expected burden is an expected repair count, and restricted survival is expected operating time before terminal failure within the horizon. This change of interpretation leaves the probability calculation unchanged. A held-out Brier score for terminal failure is reported with the same censoring convention. Real engineering deployment would require a scientifically adequate state, observation model and validation design; generic synthetic labels do not supply those requirements.

\subsection{Colorectal readmission cohort}
The \code{readmission} data distributed by \pkg{frailtypack} contain postoperative rehospitalization intervals and a terminal death indicator. \code{cs\_readmission()} preserves the recurrent episodes and uses the first recorded covariate values for its baseline example; later Charlson values remain available as recorded fields rather than being moved backward in time. Its targets are expected future admissions, death risk and restricted survival. Readmission is a utilization endpoint, not a collection of distinct disease-type marks. The held-out metrics are descriptive internal validation. The public dataset's observation scheme and limited covariates do not establish complete capture of care outside the observed system.

The new frailty sensitivity fit uses one common readmission design for all three families (Table~\ref{tab:frailty}). All three reported fits met their regularity diagnostics. The fitted lognormal model had the smallest AIC in this analysis. The shared zero-variance score statistic was approximately 29.66, with a one-sided boundary-reference $p$ value of $2.57\times10^{-8}$. This is evidence against the specified no-frailty null, not a test that the preferred mixing distribution is true. The analysis is separate from the predecessor example and retains its own session and fitted objects.

The lognormal comparison uses the refined fit, not its preliminary 40-node optimizer result. Doubling from 320 to 640 nodes changed individual log likelihoods by at most $5.90\times10^{-10}$ and mean score components by at most $2.00\times10^{-11}$. Earlier refinements failed the declared tolerance and remain in the audit table. Optimizer convergence is therefore not treated as evidence that the mixing integral has converged.

\begin{table}[tbp]\centering\small
\begin{tabular}{lrrrl}\toprule
Frailty & Log likelihood & AIC & Variance & Regular fit \\ \midrule
gamma & -870.60 & 1775.20 & 1.012 & yes \\
lognormal & -861.76 & 1757.51 & 1.827 & yes \\
discrete & -881.08 & 1798.16 & 0.929 & yes \\
\bottomrule\end{tabular}
\caption{Revision readmission sensitivity analysis on 403 subjects under a common design and observation scheme. The discrete model has two estimated mass points. AIC compares the fitted working distributions descriptively; no nonnested chi-square test is used.}\label{tab:frailty}
\end{table}

\subsection{Colon trial with day-coarsened event times}
The \code{colon} data in \pkg{survival} provide paired recurrence and death outcomes from a randomized adjuvant-treatment trial. The adapter retains all participants and treats recurrence as a non-repeatable mark. The revised automatic generator and the specialized adapter integrate a same-day recurrence and death over the legal region $d-1<t_R<t_D\leq d$, where $d$ is the recorded day. This convention preserves both events; it does not recover their actual within-day times. Integration intervals are split at baseline discontinuities. The application compares the binary \code{node4} covariate with continuous \code{nodes}, with missing-value handling stated in the adapter metadata, and checks quadrature refinement.

The trial's randomized origin does not make these fitted prognostic contrasts causal estimates. The likelihood and forecast still depend on baseline, frailty, observation and eligibility specifications. A forecast at an earlier landmark uses only a pre-landmark fibre. The full-follow-up quadrature paths used to fit the model cannot be reused as if they were already observed at that earlier time. Table~\ref{tab:validation} and the supplementary output files report validation, convergence and sensitivity diagnostics without excluding difficult fits from their denominators.

\begin{table}[tbp]
\centering\small
\begin{tabular}{lrrrrrr}
\toprule
Case & Eligible & Events & Brier & AUROC & CITL & Slope \\
\midrule
Perinatal (synthetic) & 31 & 3 & 0.114 & 0.379 & -0.611 & -2.058 \\
Oncology (synthetic) & 30 & 6 & 0.135 & 0.701 & -0.196 & 1.811 \\
Equipment (synthetic) & 28 & 2 & 0.075 & 0.557 & -0.016 & -1.401 \\
Readmission & 65 & 6 & 0.066 & 0.743 & 0.170 & 0.733 \\
Colon trial & 173 & 21 & 0.069 & 0.871 & -0.176 & 1.278 \\
\bottomrule
\end{tabular}
\caption{Held-out terminal-risk validation. Every fifth sorted subject identifier is held out; remaining subjects train the model. These small deterministic splits illustrate the workflow and are not external validation. The synthetic perinatal discrimination result is retained despite its low AUROC.}\label{tab:validation}
\begin{minipage}{\textwidth}\footnotesize CITL denotes calibration-in-the-large. Forecast duration is one stated time unit; landmarks and censoring assumptions are recorded in the accompanying metrics file. Censoring weights use the marginal Kaplan--Meier estimate. Few observed events make discrimination and calibration imprecise.\end{minipage}
\end{table}

\Needspace{12\baselineskip}
\subsection{A baseline-treatment causal workflow}\label{sec:causal_example}
The companion \code{causal/run\_causal.R} generates 1,600 synthetic subjects with one observed binary baseline confounder, two recurrent event types, one absorbing event and right censoring. Baseline treatment probability is 0.3 or 0.7 according to the confounder. Event intensities depend on treatment, confounder and retained recurrence counts. Treatment and event generation are explicit; consistency and exchangeability hold by construction. This is an implementation example, not an empirical intervention study.
\begin{CodeChunk}
\begin{CodeInput}
R> causal_fit <- cs_causal(~x, d, treatment = "treatment",
+    horizon = 1, censor = "censored", time_unit = "years",
+    event_types = c("1", "2"),
+    spec = list(history_cap = 1L, grid = 64L), seed = 81)
R> cs_causal_contrast(causal_fit)
R> cs_causal_contrast(causal_fit, "ratio")
R> cs_audit_report(causal_fit, format = "json")
\end{CodeInput}
\end{CodeChunk}
Here \code{d} is the generated start--stop dataset, \code{x} the observed confounder and \code{treatment} the fixed baseline assignment. The function estimates treatment probabilities within training folds. A known randomized allocation can instead be declared with \code{propensity = 0.5}. Terminal events stop both recurrent processes. Baseline confounders and the treatment column are explicit inputs, not inferred from the outcome marks.
\begin{table}[tbp]\centering\small
\begin{tabular}{lrrrrr}\toprule
Target & Arm 0 & Arm 1 & Difference & Pointwise 95\% CI & Oracle difference\\\midrule
Type-1 burden & 0.7088 & 0.5659 & -0.1429 & [-0.2423, -0.0435] & -0.1989 \\
Type-2 burden & 0.3113 & 0.3471 & 0.0358 & [-0.0368, 0.1084] & 0.0517 \\
Terminal risk & 0.4078 & 0.3487 & -0.0591 & [-0.1157, -0.0024] & -0.0777 \\
Time alive (years) & 0.7922 & 0.8173 & 0.0250 & [-0.0095, 0.0596] & 0.0439 \\
\bottomrule\end{tabular}
\caption{Synthetic baseline-treatment example. Both arms are standardized over the same cohort. Intervals are influence-function normal approximations. The population oracle averages equally over the two confounder strata and is calculated independently by a dense matrix exponential; sampling variation in both baseline composition and outcomes remains in the estimates.}\label{tab:causal}
\end{table}
The maximum absolute mean score is $1.84\times10^{-9}$, and doubling the time grid from 64 to 128 changes a contrast by at most $1.36\times10^{-8}$. A separate analytic exponential example agrees with the treatment-weighted native EIF to $4.37\times10^{-13}$; setting treatment probability to one reproduces the unchanged prognostic kernel exactly. These are numerical checks, not sampling-coverage guarantees.

The supplied remainder check independently simulates one million subjects in each of four nuisance configurations, sharing one generated dataset. When both event and weighting nuisances are misspecified, the observed bias is 0.06987 versus the analytic product remainder 0.07020 (Monte Carlo standard error 0.00044). The all-correct, event-correct and weight-product-correct discrepancies are each below one Monte Carlo standard error. All configurations and numerical outputs are retained; this verifies the stated identity in that design without claiming universal empirical efficiency.

\subsection{Python access to the R implementation}\label{sec:python}
The package maintains one statistical implementation. The supplied \proglang{Python} examples call the installed \proglang{R} package; they do not reimplement its likelihood, targeting or forecasting algorithms. Both routes therefore require a compatible \proglang{R} installation and \pkg{coherentSurv}. The process-based runner invokes \code{Rscript} and returns a forecast CSV and JSON audit report. The \pkg{rpy2} route retains fitted objects in the embedded \proglang{R} session and converts input and output data frames only at the interface boundary. CSV identifiers preserve leading zeros and literal \code{NA} labels. The example rejects empty inputs and missing schema columns before writing results. Stage timings and a parity-status sidecar distinguish ordinary execution from the optional verification run. Its platform and version requirements are documented at \url{https://rpy2.github.io/doc/latest/html/overview.html}.

The following calls illustrate the interactive route after a study-specific history table has been converted to an \proglang{R} data frame. The supplied executable example additionally defines the input schema, converts a pandas data frame with a local conversion context, exports diagnostics and optionally compares the result with direct \proglang{R} calls using \code{--verify-parity}.
\begin{Code}
from rpy2.robjects import Formula, StrVector, FloatVector
from rpy2.robjects.packages import importr
engine = importr("coherentSurv")
stats = importr("stats")
fit = engine.cs_fit(Formula("~ x"), data=r_data,
                    event_types=StrVector(["1", "2"]),
                    time_unit="months")
forecast = stats.predict(fit, newdata=r_history, landmark=0,
                         horizon=FloatVector([.25, .5, 1]),
                         type="all")
\end{Code}
Here \code{r\_data} contains training interval histories, \code{r\_history} supplies the histories to forecast, \code{x} is a measured covariate, event labels \code{1} and \code{2} denote the two recurrent marks, and all durations are in months. The landmark and future durations specify the prediction question; the example values are not universal defaults for a clinical application. The ordinary runner fits once; the optional verification mode performs a second fit solely to test transport equivalence. Conversion can copy data and carries a computational cost. Retaining the fitted object and passing complete data batches avoids repeated model fitting and per-row transfers. This access layer does not itself establish compatibility with scikit-learn pipelines or add inferential guarantees.

The naming convention uses the \code{cs\_} prefix for package-specific functions, including \code{cs\_fit()}, \code{cs\_control()}, \code{cs\_as\_trajectory()} and \code{cs\_build\_history()}. Standard \proglang{R} generics retain their familiar names. This release changes the development API; earlier serialized model objects should be refitted using the current package. Python-interface validation is recorded separately from the statistical benchmarks, with the executed platforms, dependency versions and maximum differences from direct \proglang{R} calls.

\section{Validation and reproducibility}\label{sec:reproducibility}
A related survey-plasmode study of rare cardiovascular--kidney--metabolic multimorbidity used whole-primary-sampling-unit cross-validation, fold-contained preprocessing and cross-fitted recalibration to evaluate performance under complex sampling \citep{AdetunjiSurvey2026}; here, the validation unit and observation assumptions must likewise be specified for the event-history analysis.

\code{validate()} evaluates supplied held-out risks at an explicit horizon. Cases have a terminal event by the horizon and controls remain observed beyond it. Inverse probabilities of censoring weight the Brier score, AUROC and threshold table. The default Kaplan--Meier censoring estimate requires marginal independent censoring; analysts may instead supply justified conditional censoring probabilities. The function does not estimate such a conditional model automatically. Follow-up times must be measured from the prediction landmark after restricting to eligible subjects; unused arguments are rejected so that an unsupported landmark argument cannot silently change the intended validation question.

Calibration-in-the-large is estimated as a logistic intercept with predicted log odds as an offset; a separate slope summarizes their multiplier. The integrated calibration index uses a weighted smooth of observed outcome against predicted risk. Sensitivity, specificity, positive and negative predictive values, and net benefit share a threshold-specific weighted table. Undefined quantities remain labelled rather than being replaced by zeros. Complete or quasi separation is reported explicitly and the calibration slope is withheld when a finite logistic maximum-likelihood estimate does not exist. Ordinary model summaries withhold Wald limits when the fit is flagged as nonregular. These statistics quantify different properties: AUROC does not establish calibration, a Brier improvement does not establish clinical benefit, and an internal held-out estimate does not replace external validation.

The companion \code{code.R} entry point reconstructs the manuscript tables and figures from package calculations and the supplied recorded experiment outputs. Its default mode reproduces the article from these outputs; separate fresh-fit and experiment scripts rerun the underlying analyses. It also records session information, source hashes and expected outputs. Runtime and memory are inherently machine dependent; deterministic numerical outputs and tolerances are checked separately. The source distribution includes tutorials, help pages, structured errors and a test suite. Revision checks record their own version and runtime rather than inheriting a predecessor check result. Standalone C++ tests use strict compiler warnings. The supplied check records identify the source revision, platform, compiler and executed tests.

Version 0.11.1 adds validation-argument checks, separation-aware calibration reporting, guarded coefficient intervals through summary, table and \code{confint()} methods, and a single-fit Python runner. It also adds explicitly scoped source mappings, environment diagnostics and code-specific error conditions. Its verification is recorded separately from the 0.11.0 naming revision. Earlier benchmark runtimes and simulation results retain their original version provenance; they are not relabelled as newly repeated experiments.

Version 0.11.1 passed a full local \code{R CMD check --as-cran}
under Windows \proglang{R}-devel with remote incoming checks enabled:
0 errors, 0 warnings and one incoming-feasibility note identifying the
maintainer and new submission. All 1,476 test expectations, ten vignettes,
examples and PDF/HTML manual checks passed. Seven GitHub platform jobs
passed: Linux, macOS and Windows under \proglang{R}-release and
\proglang{R}-devel, and Linux under \proglang{R}-oldrel. The standalone
native fixtures passed four GCC/Clang configurations, including
AddressSanitizer and UndefinedBehaviorSanitizer. These checks include
the causal kernel. All 23 native source and header files in the source
archive agree with 0.11.0 after line-ending normalization.

The direct \pkg{rpy2} example was executed on Linux with \proglang{R}
4.6.2, \pkg{rpy2} 3.6.9 and pandas 2.3.3. It returned 15 forecast rows
and matched direct \proglang{R} calls exactly for coefficients, covariance
and forecasts. Identifier preservation and invalid-input checks passed.
The \code{Rscript} example was executed separately on Windows. All 50
companion reporting checks passed. The manuscript replication runner
was executed using the retained experiment outputs; the original Monte
Carlo studies were not rerun for this interface revision. These checks
establish agreement for the tested calculations, not external predictive
validity, regulatory qualification or CRAN acceptance. The source commit,
archive checksums, environments and executed logs accompany the revision.

The replication materials provide software source, commented scripts, recorded outputs, seeds and session information \citep{JSS}. Source code is distributed under GPL-3 in the public repository \url{https://github.com/setrojalab/coherentSurv}. Installation instructions and documentation are available at \url{https://setrojalab.github.io/coherentSurv/}. The 0.11.1 review candidate and its verification records are linked from \url{https://github.com/setrojalab/coherentSurv/pull/4}; the published 0.11.0 release remains separately available. The revision includes the source archive, reference manual, article sources and computational supplement. The renamed source archive is prepared for CRAN review as the successor to the pending submission under the earlier package name. CRAN acceptance is not implied by the availability of a source archive. The same materials accompany the journal submission.

For installation from the accompanying 0.11.1 source archive, use:
\begin{Code}
install.packages("coherentSurv_0.11.1.tar.gz",
                 repos = NULL, type = "source")
\end{Code}
Source installation requires the declared R dependencies and a C++17 toolchain; Windows users require the Rtools version appropriate to their R installation. Installation of the supplied source archive also provides ten prebuilt vignettes.

\subsection{Declared source mappings and installation diagnostics}
The \code{cs\_from\_adam()} helper selects one ADTTE parameter and maps explicitly declared event and censor codes, preserving the supplied duration definition and optional ADSL covariates. The \code{cs\_from\_omop()} helper accepts a reviewed event extract, a supplied concept map and one continuous observation period per person under an explicitly declared exact-day-boundary convention. Duplicate or simultaneous events, unsupported entry histories and inconsistent containment raise errors. Neither helper certifies the source standard or infers complete ascertainment from the presence of a database table. Date-coarsened clinical histories still require the observation model appropriate to their timing information.

The offline \code{cs\_validate\_env()} diagnostic records loaded software metadata and executes four small analytic checks: a Poisson-boundary likelihood, dense and sparse absorbing-state propagation, and Louis mixture information. Results distinguish passed, failed and unavailable information. In particular, a compiler currently installed on a machine does not identify the compiler used to build an existing binary. Optional elapsed times and returned-object sizes describe these fixtures only; they are not peak-memory measurements or comparative performance benchmarks. Reports can be exported as JSON or text, without treating an unsigned diagnostic as a qualification certificate.

\subsection{Audit export, offline execution and study governance}
\code{cs\_audit\_report()} exports a list, JSON or text report from a fitted object, optionally summarizing numerical diagnostics from predictions and hashing explicitly supplied source files. SHA-256 fingerprints of version-3 R serialization identify the captured data object; source-file fingerprints identify file bytes. Capture-time/session metadata and report-export metadata are distinguished. Legacy MD5 fingerprints remain labelled for compatibility. A hash establishes an integrity comparison for its stated representation, not data anonymity, a digital signature, immutable storage or regulatory approval.

The model and prediction routines can run offline with a suitable installed R environment, the compiled package and its open-source dependencies. Preparing that environment and obtaining optional datasets are separate steps. Reproducible installation instructions and machine checks support an analyst's validation record; they do not certify an institution's operating environment.

ICH E9(R1) asks trial analyses to align with a defined treatment-effect estimand and its intercurrent-event strategy \citep{ICHE9R1}. The prognostic APIs do not automatically identify such effects. The separate baseline-treatment interface provides the estimands and assumptions in Section~\ref{sec:causal}; it does not select an intercurrent-event strategy for the analyst. Optional analyst declarations supplied to \code{cs\_fit()} or \code{cs\_causal()} are retained in summaries and audit exports; they do not alter the implemented targets or verify identification. A submission analysis still needs its protocol, estimand, prespecified analysis, sensitivity strategy, source traceability and fit-for-purpose validation. Schema mappings, software hashes and offline execution do not constitute regulatory qualification.

Platform status is reported from executed logs. The supplied check records distinguish the completed 0.11.0 naming-revision checks, the current 0.11.1 interface checks and the earlier method experiments.

\Needspace{10\baselineskip}
\section{Discussion and development scope}\label{sec:discussion}
\pkg{coherentSurv} provides a common interface for event-history likelihoods and forecasts while exposing separate observation experiments for targeting, sensitivity inference and prediction calibration. Its main practical benefit is traceability: an analyst can inspect which histories were admitted, which constraints define a bound, which subject contributed an influence-function value, and which numerical approximation produced a forecast. The benchmark and examples demonstrate these computations without interpreting machine precision as empirical validity.

The five engineering requests are implemented as explicit modules. Automatic fibres remove manual order enumeration for the declared complete-count timing experiment. CSR propagation removes dense transition storage, and structural optimization handles large supports for the specified recording constraints. Finite-path and continuous-time targeting are exposed in the main API. Future covariates can follow schedules or estimated Gaussian submodels. Alternative frailty laws and boundary diagnostics make distributional sensitivity inspectable.

Their statistical conditions remain consequential. An automatic timing fibre does not reconstruct an unknown number of hidden events. Sparse storage does not make an insufficient retained state Markov. Logit transformation does not guarantee sparse finite-sample coverage. Gaussian evolution requires its own measurement and feedback assumptions, and finite mass-point fits can be nonregular. These boundaries are attached to the relevant outputs so that the analyst can assess applicability rather than infer a guarantee from a function name.

Future work includes richer censoring and longitudinal nuisance models, additional observation adapters and separately verified language interfaces. Cost and quality-of-life valuation require explicit reward definitions. Static baseline treatment contrasts are implemented through \code{cs\_causal()}. Time-varying policies, preventable fractions and treatment recommendations require additional causal estimands and identification beyond that static extension. New features should retain their own observation contract, independent oracle and validation evidence.

A possible four-package organization is recorded as a development roadmap, not four released packages. \pkg{coherentSurv} would retain event-history likelihoods, common-law forecasting and the shared numerical core; \pkg{recurbounds} would concentrate compatibility sets and sensitivity inference; \pkg{recurcausal} could extend the implemented static baseline-treatment interface to separately justified longitudinal policies; and \pkg{recurmetrics} would organize validation, calibration and decision reporting. In this release these names do not split the repository or create new dependencies. The implemented sensitivity and validation interfaces remain in \pkg{coherentSurv}; the additional reporting script remains explicitly sourced companion code. Any later separation must preserve contracts, numerical equivalence, dependency direction and independent tests. The downstream package names remain proposed; the baseline-treatment causal capability described here is implemented in \pkg{coherentSurv}.

\section*{Computational details}

\paragraph{Package version.} This review candidate is \pkg{coherentSurv} version 0.11.1. The fitting and control entry points are \code{cs\_fit()} and \code{cs\_control()}; the native implementation uses a consistent package namespace and include path. The native numerical kernels and the scientific interpretation of the historical benchmarks are unchanged; the validation and reporting safeguards have separate revision tests.

The software version described here is \code{0.11.1}; this manuscript was revised on 26 November 2026. The predecessor benchmark and five-case outputs retain their original snapshot \code{0.9.0.9000} and candidate \code{0.9.1} provenance under \proglang{R} 4.4.2 UCRT and GCC 14.3 on Windows 11. The 0.10.0 module validations retain their original provenance; the targeted 0.10.1 API checks have a separate record. A C++17 core, GPL-3 source license, LaTeX sources, bibliography, executed causal vignette and generated figures accompany the materials. Recorded results are version-labelled; the original experiment outputs are retained alongside the current interface checks.

\section*{Data availability and ethics}
The three bundled example datasets are simulated and include generating parameters and seeds. The two clinical examples use data already distributed by the \pkg{frailtypack} and \pkg{survival} packages, loaded through their documented interfaces. The causal example generates an additional synthetic cohort. The additional survival-wrapper example uses \code{survival::lung}; it is a worked API example, not an external validation cohort. No new participants were recruited. The analyses use the public example datasets as distributed by their source packages; no institutional ethics determination is asserted here. Patient-level clinical records are excluded from the software repository materials.

\section*{Funding and competing interests}
The authors report no funding and no competing interests.
\bibliography{refs}
\end{document}